\documentclass[%
reprint,
superscriptaddress,
 amsmath,amssymb,
 aps,
prd,
]{revtex4-2}

\usepackage{graphicx}
\usepackage{dcolumn}
\usepackage{bm}

\begin{document}

\title{Investigating galactic double white dwarfs for sub-milliHz gravitational wave mission ASTROD-GW}


\author{Gang Wang}
\email[Gang Wang: ]{gwang@shao.ac.cn, gwanggw@gmail.com}
\affiliation{Shanghai Astronomical Observatory, Chinese Academy of Sciences, Shanghai, 200030, China}

\author{Zhen Yan}
\affiliation{Shanghai Astronomical Observatory, Chinese Academy of Sciences, Shanghai, 200030, China}

\author{Bin Hu}
\affiliation{Institute for Frontier in Astronomy and Astrophysics, Beijing Normal University, Beijing, 102206, China}
\affiliation{Department of Astronomy, Beijing Normal University, Beijing 100875, China}

\author{Wei-Tou Ni}
\email[Wei-Tou Ni: ]{weitou@gmail.com}
\affiliation{State Key Laboratory of Magnetic Resonance and Atomic and Molecular Physics, Innovation Academy for Precision Measurement Science and Technology (APM), Chinese Academy of Sciences, Wuhan 430071, China}
\affiliation{Department of Physics, National Tsing Hua University, Hsinchu, Taiwan, 30013, ROC}

\date{\today} 

\begin{abstract}

A large number of galactic binary systems emit gravitational waves (GW) continuously with frequencies below $\sim$10 mHz. The LISA mission could identify tens of thousands of binaries over years of observation and will be subject to the confusion noise around 1 mHz yielded by the unresolved sources. Beyond LISA, there are several missions have been proposed to observe GWs in the sub-mHz range where the galactic foreground is expected to be overwhelming the instrumental noises. In this study, we investigate the detectability of sub-mHz GW missions to detect the galactic double white dwarf (DWD) binaries and evaluate the confusion noise produced by the undistinguished DWDs. This confusion noise could also be viewed as a stochastic GW foreground and be effectively observed in the sub-mHz band. The parameter determinations for the modeled foreground are examined by employing different detector sensitivities and population models. By assuming the determined foregrounds could be subtracted from the data, we evaluate the residuals which are expected to have power spectral densities two orders of magnitude lower than the originals data.

\end{abstract}

\maketitle

\section{Introduction}

During the O1-O3 runs, dozens of gravitational waves (GWs) events from compact binary coalescences have been identified by advanced LIGO and advanced Virgo \cite[and references therein]{LIGOScientific:2018mvr,LIGOScientific:2020ibl,LIGOScientific:2021djp,Nitz:3OGC}. The KAGRA will join the O4 run with advanced LIGO and advanced Virgo in 2023 \footnote{https://www.ligo.caltech.edu/page/observing-plans}. These ground-based laser interferometers, as well as the planned Einstein Telescope \cite{Punturo:2010zz} and Cosmic Explorer \cite{Reitze:2019iox}, observe the GWs in the high-frequency band from few-Hz to kilo-Hz.

The BBO \cite{BBO:2005} and DECIGO \cite{DECIGO:2006} were first proposed to observe the GW in deci-Hz band in space. The recently proposed space-based detectors, for instance, DO \cite{DO:2019}, AMIGO (Astrodynamical Middle-frequency Interferometric Gravitational wave Observatory, to differ from the following AMIGO) \cite{Ni:2017bzv,Ni:2019nau}, AIGSO \cite{Gao:2017rgh,Wang:2019oeu}, and AGIS-MAGIS \cite{Hogan:2011tsw,Hogan:2015xla,Graham:2017pmn}, AEDGE \cite{AEDGE:2019nxb}, are also targeting to observe the GW in the deci-Hz frequency band. Besides the space missions, ground-based detectors are also proposed by employing various technologies including AION \cite{AION:2019}, ELGAR \cite{ELGAR:2019}, MIGA \cite{MIGA:2017}, ZAIGA \cite{ZAIGA:2019}, SOGRO \cite{SOGRO:2016}, TOBA \cite{TOBA:2018}, etc. (see, e.g. \citep{Gao:2021esp} for various recent activities in the middle-frequency band).

The space-borne missions, LISA \cite{LISA2017}, Taiji \cite{Hu:2017mde}, and TianQin \cite{TianQin:2015yph} are actively developing and scheduled to be launched in the 2030s. These missions will observe the GW in the mHz low-frequency band (0.1 mHz–0.1 Hz). With the success of the LISA Pathfinder, the drag-free technology, as one of the core requirements, is verified for the LISA mission \cite[and references therein]{LPF:2016,Armano:2018kix}. The next-generation mission, Advanced Millihertz Gravitational-wave Observatory (AMIGO, to differ from aforesaid AMIGO) \cite{Baibhav:2019rsa}, has been proposed with a sensitivity better than the LISA by one order.

Beyond the mHz low-frequency range, missions are proposed to observe the GW in sub-milli-Hz (sub-mHz) or even lower frequencies which requires a longer interferometric arm than LISA. The mission concepts, ASTROD-GW \cite[and references therein]{Ni:2012eh,Ni:2016wcv}, Folkner mission \cite{FolknerMission:2012,FolknerMission} and LISAmax \cite{Martens:2023mgm} are designed to employ three spacecraft in a heliocentric orbit to form a triangular interferometer with an arm length of $\sqrt{3}$ AU (as shown in Fig. \ref{fig:SC_layout}). The $\mu$Ares will utilize an interferometer in a Martian orbit to form a longer arm \cite{muAres:2019}, and the Super-ASTROD is a more ambitious mission and designed to place spacecraft in the Jovian orbit \cite{Ni:2008bj}. For ASTROD-GW or Folkner mission, the interferometric arm could be $\sim$100 times longer than LISA's $2.5 \times 10^6$ km arm length, and their sensitive frequencies could be two orders lower than the LISA.

The mission to detect the GW in low frequencies would be subject to the influence of galactic sources. Enormous galactic binaries are in the early inspiral stage, and they will continuously emit the GW with frequencies lower than $\sim$10 mHz. The LISA mission could identify thousands to tens of thousands of galactic binaries in years of observation \cite[and references therein]{Vecchio:2004ec,Nissanke:2012eh,Cornish:2017vip,Korol:2017qcx,Amaro-Seoane:2022rxf}, and some binaries in the neighboring galaxy may also be detected by the LISA \cite{Korol:2018ulo,Korol:2020lpq}. The GW signals from unresolved sources would overlap and become a confusion noise, and this noise would affect the LISA's sensitivity around 1 mHz \cite[and references therein]{Nelemans:2001hp,Timpano:2005gm,Ruiter:2007xx,Lamberts:2019nyk}. The identifications and subtractions of galactic binaries have been performed numerically for the LISA(-like) missions \cite{Cornish:2003vj,Zhang:2021htc,Zhang:2022wcp,Liu:2023qap}, and the global analysis is developing to resolve different GW sources simultaneously \cite{Littenberg:2020bxy,Littenberg:2023xpl}. On the other side, the confusion noise may be characterized and/or mitigated by post-processing \cite{Karnesis:2021tsh,Digman:2022jmp,Lin:2022huh}. Since GWs from galactic binaries are (nearly) monochromatic and last for years, the efficient calculation of time-varying antenna patterns could be required to achieve accurate parameters inference, and a solution for the efficient analysis is developed by utilizing the graphics processing unit (GPU) \cite{Katz:2022yqe}.

Compared to the galactic population in the mHz band, much more binaries emit the GW in sub-mHz frequencies, and the missions in this band will face severer foreground. The spectrum of confusion noise could overwhelm the instrumental noise by orders as notified in \cite{FolknerMission,muAres:2019}. The observable population will depend on the astrophysical processes of binary formation and evolution. In this work, we employ four DWD populations simulated in \citep{Thiele:2021yyb,thiele_data} by using the population synthesis suite \texttt{COSMIC} \cite{Breivik:2019lmt,cosmic_software}. The ASTROD-GW with two different sensitivity (elementary and advanced) configurations are utilized to investigate their detectability to the galactic DWDs. And the confusion noises yielded by the unresolved sources are evaluated and characterized. Furthermore, the confusion noises are modeled by implementing a polynomial fitting. The confusion noises are simulated as a stochastic GW foreground and injected into the simulated data, and the parameter determinations of foregrounds are performed by using the mock data. By optimistically assuming the galactic foreground could be subtracted from the data, the residuals after the subtractions are evaluated for different sensitivity configurations and population models, and the power spectral densities of the residuals would be two orders lower than the original levels.

This paper is organized as follows. 
In Sec. \ref{sec:missions}, we introduce the mission configurations and the corresponding sensitivities from the first-generation Michelson time-delay interferometry.
In Sec. \ref{sec:galactic_binaries}, we specify the galactic DWD populations and the identifications of the mission to these sources. And then the confusion noises from unresolved binaries are evaluated and modeled with a polynomial fitting.
And in Sec. \ref{sec:characterization}, the parameter estimations for the modeled foregrounds are performed by using the simulated data, and the corresponding residuals are evaluated by subtracting the restored foregrounds from the data spectra.
We recapitulate our conclusions and discussions in Sec. \ref{sec:conclusions}. (We set $G=c=1$ in this work except otherwise stated).

\section{sub-milliHz missions} \label{sec:missions}

\subsection{Mission orbit} \label{subsec:orbit}

\begin{figure}[htb]
\includegraphics[width=0.45\textwidth]{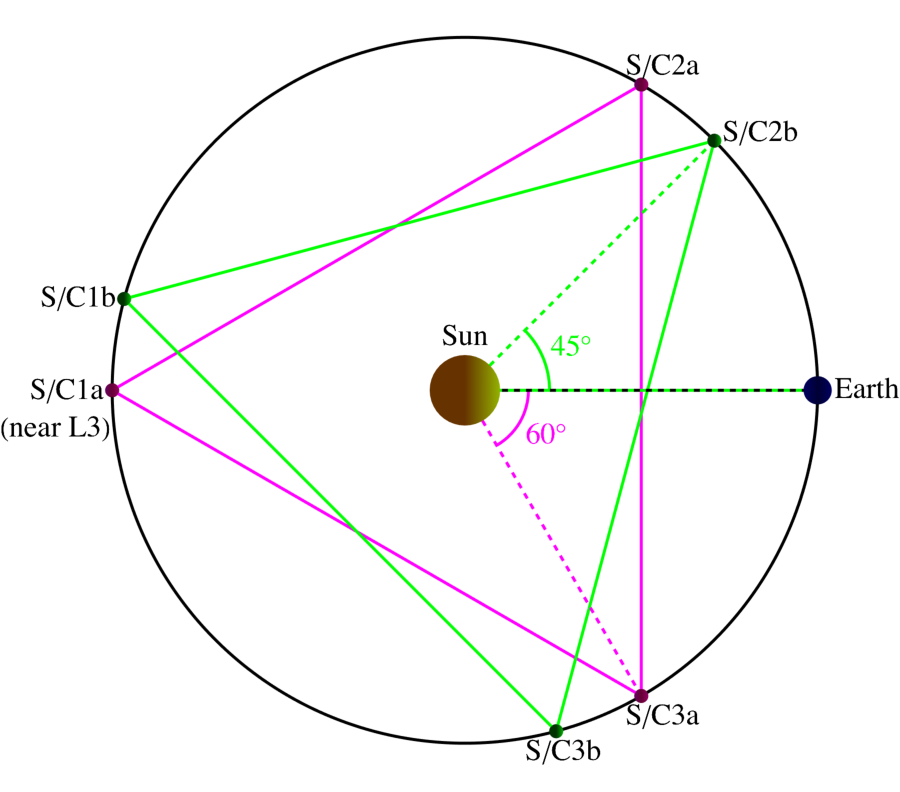}
\includegraphics[width=0.45\textwidth]{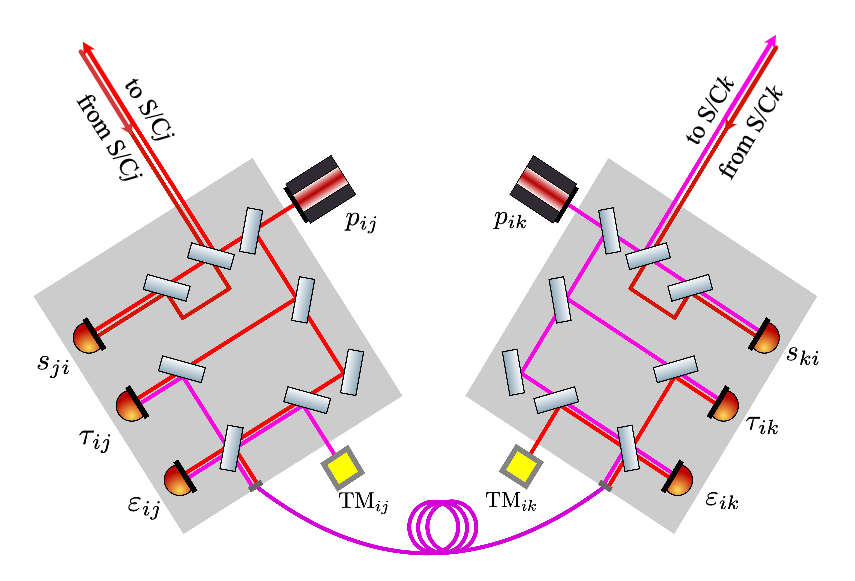} 
\caption{The constellation formations of ASTROD-GW mission (magenta) and Folkner mission (green) \cite{Ni:2012eh,FolknerMission}. For each mission, three spacecraft are numbered in the clockwise direction. The ASTROD-GW is designed to deploy three spacecraft near the Sun-Earth Lagrange points $L_3, L_4$ and $L_5$, and the nearest spacecraft of Folkner mssion is leading the Earth by 45$^\circ$. The diagram of two optical benches on S/C$i$ for ASTROD-GW (lower panel) which is treated as same as the LISA \cite{Otto:2015,Wang:2022sti}. \label{fig:SC_layout}
}
\end{figure}

Missions have been proposed to observe GWs in the sub-mHz band. The ASTROD-GW is proposed in 2009 and designed to deploy three spacecraft (S/C) near the Sun-Earth Lagrange points $L_3, L_4$ and $L_5$ \cite{ni2009astrod,ni2010astrod,ni2009sanya,Ni:2011ib,Men2010,Wang:2012fqs}, as illustrated in Fig. \ref{fig:SC_layout}. Three S/C could form a triangular interferometer with an arm length of $2.6 \times 10^8$ km (1.732 AU $\simeq$ 866.67 s). To improve the antenna pattern, alternative orbits are designed to make the S/C formation plane precession with mission time \cite{Wang:2014uaw}. The orbital configuration of the Folkner mission was initially proposed during the GW Mission Concept Study in 2012 which choose a formation similar to the ASTROD-GW \cite{FolknerMission:2012}, and a more explicit mission concept is announced with an altered configuration in 2019 \cite{FolknerMission}. Three S/C are also deployed on the heliocentric Earth-like orbit, and the nearest S/C is leading the Earth by 45$^\circ$ as shown in Fig. \ref{fig:SC_layout}. Compared to ASTROD-GW, the Folkner mission places one S/C at a closer distance from Earth which could reduce the difficulty of communication with the constellation. The $\mu$Ares is designed to place three S/C in a Martian orbit and form an equilateral triangle \cite{muAres:2019}, and it could observe even lower frequency GWs than ASTROD-GW and Folker mission.

\begin{figure}[htb]
\includegraphics[width=0.46\textwidth]{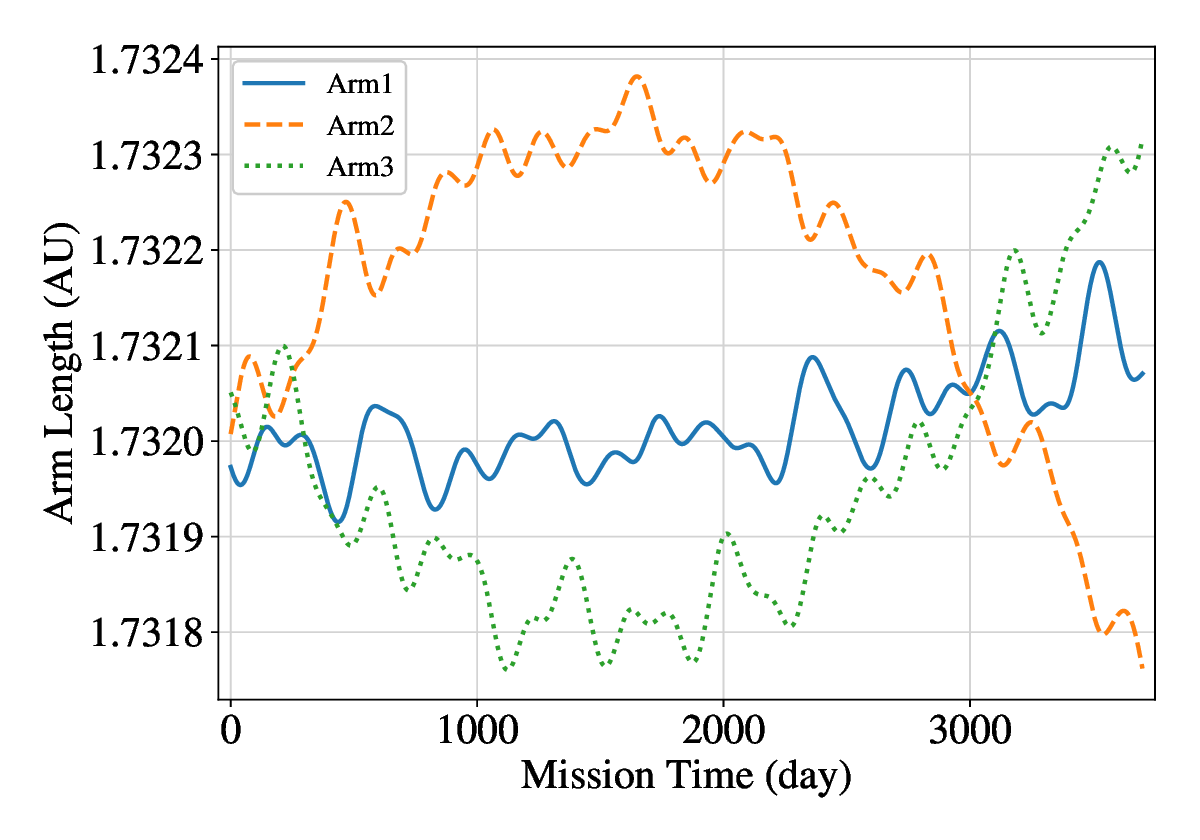}
\includegraphics[width=0.46\textwidth]{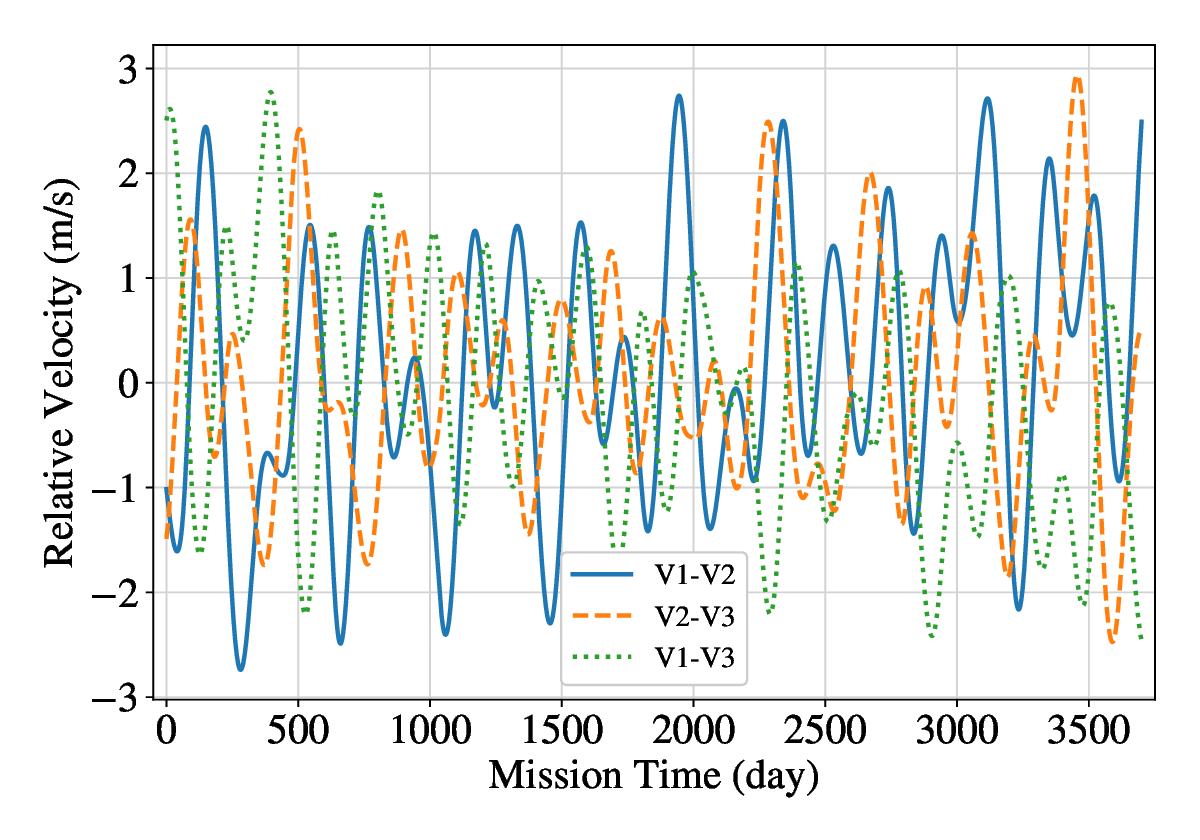} 
\includegraphics[width=0.46\textwidth]{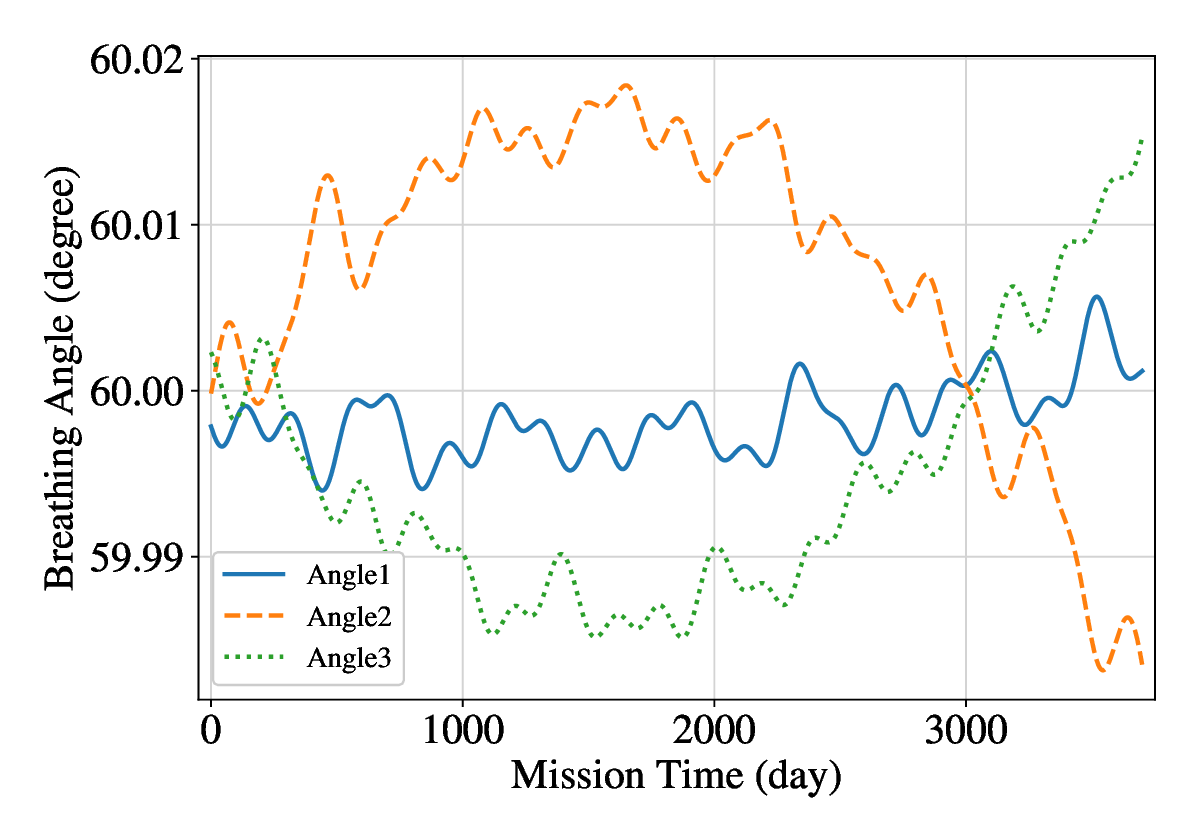} 
\caption{The arm lengths (upper), relative velocities between spacecraft (middle), and breathing angles (lower) of ASTROD-GW in 10 years \cite{Wang:2014uaw}. The lengths of three arms in a range of $1.7321 \pm 0.00035 $ AU, and arm difference less than $\sim$ 0.04\%. The amplitudes of Doppler velocities between the S/C are less than 3 m/s, and the variances of breathing angles are less than $0.02^\circ$. \label{fig:ASTROD_GW_orbit}
}
\end{figure}

Compared to the LISA orbital formation, the constellations (nearly) in the ecliptic plane could be more stable, and the arms of the interferometer could be closer to equilateral. A numerical orbit for the ASTROD-GW in 10 years is shown in Fig. \ref{fig:ASTROD_GW_orbit} which is obtained by employing an ephemeris framework \cite{Wang:2014uaw}. The lengths of three arms in a range of $1.7321 \pm 0.0004 $ AU, and arm difference less than $\sim$0.04\%. The amplitudes of Doppler velocities between the S/C are less than 3 m/s, and the variances of breathing angles are less than $0.02^\circ$. For the LISA mission, the Doppler velocities between S/C are required to be within $\pm$ 5 m/s, and the breathing angle should be in $\pm 1^\circ$ \cite{LISA2017}.

Being beneficial from the stability of Sun-Earth Lagrange points, the mission orbit of ASTROD-GW could remain stable for even more than 10 years \cite{Men2010}. The Folkner mission or $\mu$Ares, by choosing a planet-like orbit, may also be stable for a decade. A stable orbit and more equal interferometric arms would be helpful to implement the time-delay interferometry (TDI) which is developed to reduce the laser frequency noise due to the optical path differences. The ASTROD-GW is selected to perform the investigation of galactic DWD observations in the sub-mHz band. 

\subsection{Time-delay interferometry} \label{subsec:TDI}

TDI is employed by space-borne GW interferometers to suppress laser frequency noise, and it essentially synthesizes the time-shifted measurement links to construct equivalent equal paths. As the fiducial case, the Michelson is elected from five first-generation TDI configurations to evaluate the sensitivity of the mission. Three channels from Michelson are named X, Y, and Z depending on the starting/ending spacecraft \cite{1999ApJ...527..814A,2000PhRvD..62d2002E,Tinto:2020fcc}. The Michelson-X could be expressed as
\begin{equation}
\begin{aligned}
{\rm X} =& ( \mathcal{D}_{31} \mathcal{D}_{13} \mathcal{D}_{21} \eta_{12}  + \mathcal{D}_{31}  \mathcal{D}_{13} \eta_{21}  +  \mathcal{D}_{31} \eta_{13} +  \eta_{31}   ) \\ 
 & - ( \eta_{21} + \mathcal{D}_{21} \eta_{12} +\mathcal{D}_{21} \mathcal{D}_{12} \eta_{31} + \mathcal{D}_{21}  \mathcal{D}_{12} \mathcal{D}_{31} \eta_{13} ), \label{eq:X_expression}
\end{aligned}
\end{equation}
where $\mathcal{D}_{ij}$ is a time-delay operator, $ \mathcal{D}_{ij} \eta(t) = \eta(t - L_{ij} )$, $L_{ij}$ is the arm length from S/C$i$ to $j$, $\eta_{ji}$ are Doppler measurements from S/C$j$ to S/C$i$. The payload designs for the ASTROD-GW are assumed to be the same as the current LISA as specified in \cite{Otto:2012dk,Otto:2015} and illustrated in Fig. \ref{fig:SC_layout}. Two optical benches are deployed on each S/C, and three interferometers are installed on each optical bench which are science interferometer $s_{ji}$, test mass interferometer $\varepsilon_{ij}$, and reference interferometer $\tau_{ij}$. The measures are different depending on the arrangements of optical benches. For the receiving optical benches in the counterclockwise direction (S/C2$\rightarrow$S/C1, S/C3$\rightarrow$S/C2, S/C1$\rightarrow$S/C3), the interferometer measurements will be
\begin{equation} \label{eq:s_epsilon_tau_1}
\begin{aligned}
   s_{ji} = & y^h_{ji}:h + \mathcal{D}_{ji} p_{ji}(t) - p_{ij}(t) + n^{\rm op}_{ij}(t), \\
   \varepsilon_{ij} = & p_{ik}(t) - p_{ij}(t) + 2 n^{\rm acc}_{ij}(t) , \\
   \tau_{ij} = & p_{ik}(t) - p_{ij}(t) ,
\end{aligned}
\end{equation}
and measurements on receiving optical benches in the clockwise directions (1$\rightarrow$2, 2$\rightarrow$3 and 3$\rightarrow$1) will be
\begin{equation} \label{eq:s_epsilon_tau_2}
\begin{aligned}
   s_{ji} &=  y^h_{ji}:h + \mathcal{D}_{ji} p_{ji}(t) - p_{ij}(t) + n^{\rm op}_{ij}(t), \\
   \varepsilon_{ij} &= p_{ik}(t) - p_{ij}(t) - 2 n^{\rm acc}_{ij}(t) , \\
   \tau_{ij} &= p_{ik}(t) - p_{ij}(t),
\end{aligned}
\end{equation}
where $y^h_{ji}$ is the response function to a GW signal $h$ (see specific formula in Appendix \ref{sec:appendix_response}) \cite{1975GReGr...6..439E,1987GReGr..19.1101W,Vallisneri:2007xa,Vallisneri:2012np}, $p_{ij}$ is laser noise on the optical bench of S/C$i$ pointing to S/C$j$, $n^{\mathrm{op}}_{ij}$ denotes the optical path noise on the S/C$i$ pointing to $j$, and $n^{\mathrm{acc}}_{ij}$ is the acceleration noise from test mass on the S/C$i$ pointing to $j$. A measurement between two S/C is composed from multiple interferometers,
 \begin{equation} \label{eq:eta}
\begin{aligned}
  \eta_{ji} &= s_{ji} + \frac{1}{2} \left[ \tau_{ij} - \varepsilon_{ij} + \mathcal{D}_{ji} ( 2 \tau_{ji} - \varepsilon_{ji} - \tau_{jk} ) \right] \\
  & \quad \mathrm{for} \  (2 \rightarrow 1), (3 \rightarrow 2) \ \mathrm{and} \ (1 \rightarrow 3), \\
  \eta_{ji} &= s_{ji} + \frac{1}{2} \left[ \tau_{ij} - \varepsilon_{ij}  + \mathcal{D}_{ji} ( \tau_{ji} - \varepsilon_{ji}    ) + \tau_{ik} -  \tau_{ij} \right] \\
   & \quad \mathrm{for} \  (1 \rightarrow 2), (2 \rightarrow 3)\ \mathrm{and}\ (3 \rightarrow 1).
\end{aligned}
\end{equation}
For a space interferometer antenna with six laser links, three orthogonal channels (A, E, T) could be constructed from three regular channels (X, Y, Z) \cite{Prince:2002hp,Vallisneri:2007xa},
\begin{equation} \label{eq:optimalTDI}
 {\rm A} =  \frac{ {\rm Z} - {\rm X} }{\sqrt{2}} , \quad {\rm E} = \frac{ {\rm X} - 2 {\rm Y} + {\rm Z} }{\sqrt{6}} , \quad {\rm T} = \frac{ {\rm X} + {\rm Y} + {\rm Z} }{\sqrt{3}}.
\end{equation}
The A and E channels could observe GW effectively and are treated as science channels. The T channel is a null channel and insensitive to GW in the low-frequency band. Three optimal channels represent the eventual sensitivity of the mission.

\begin{figure}[htb]
\includegraphics[width=0.46\textwidth]{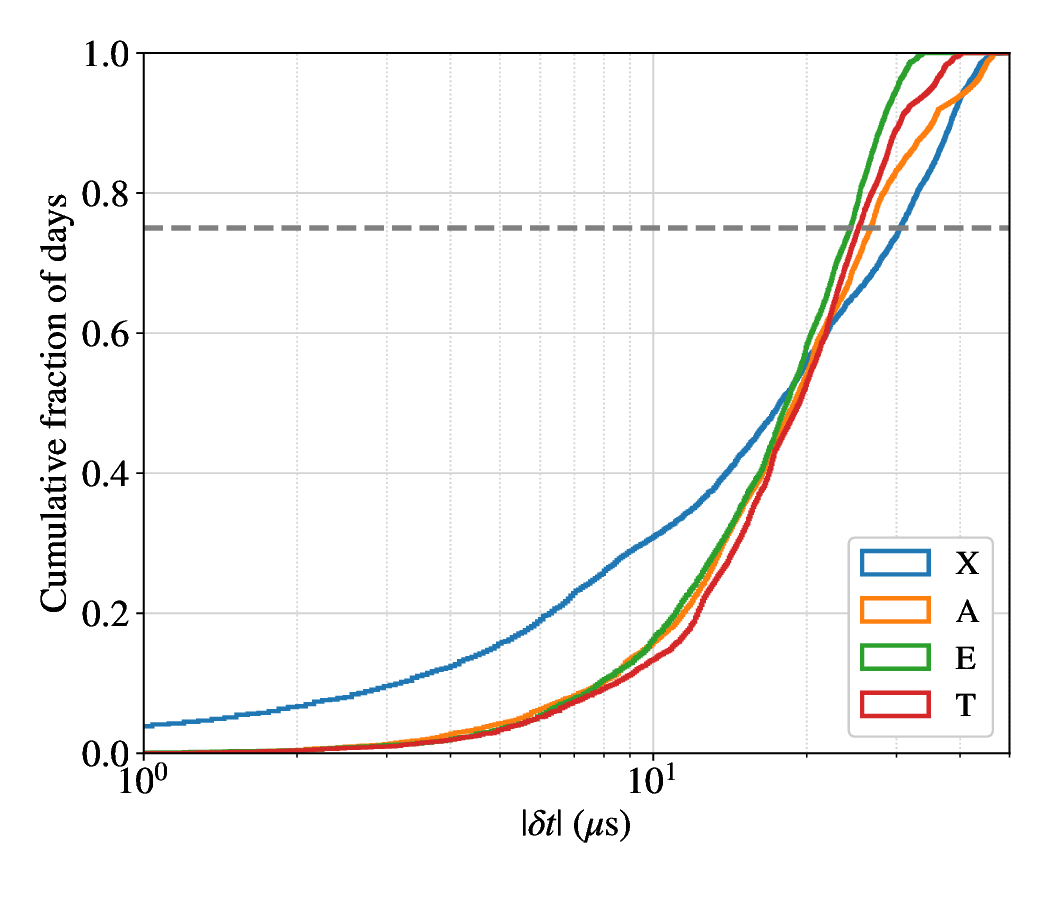}
\caption{The cumulative histograms of path mismatches in the TDI channels Michelson-X, A, E, and T for ASTROD-GW. The path mismatches are less than 5 $\mu$s in 10 years, and the mismatches are less than 3 $\mu$s in 75\% of observation time as indicated by the horizontal dashed line of 75 percentile. \label{fig:TDI_dt}
}
\end{figure}

Due to the orbital dynamics, the laser noise could not be fully canceled by the TDI. The residual laser noise is related to the optical path mismatch in a TDI channel, $\delta t_\mathrm{TDI}$,
\begin{equation}
 S_{\rm laser, TDI} \simeq ( 2 \pi f \delta t_\mathrm{TDI} )^2 S_{\rm laser},
\end{equation}
where $S_{\rm laser}$ is the power spectral density (PSD) of the laser noise. By employing the numerical algorithm in \cite{Wang:2ndTDI}, the path mismatches for the Michelson TDI channels are obtained and shown in Fig. \ref{fig:TDI_dt}. The duty cycle of the LISA mission is expected to be 75\% \cite{Seoane:2021kkk}, and we suppose this duty cycle is applicable for the sub-mHz mission(s). The 75\% of the path mismatches of the first-generation Michelson TDI channels could be less than 3 $\mu$s, and the mismatches in full period is smaller than 5 $\mu$s.
The laser source is assumed to be the same as the LISA which utilizes Nd:YAG laser with a stability of $30\ {\rm Hz}/\sqrt{\rm Hz}$, and the noise PSD will be $S_{\rm laser} = 10^{-26} / {\rm Hz} $. The PSD of laser residual noise from the least 75\% percentiles is shown by the gray area in Fig. \ref{fig:sensitivity} upper panel. The laser noise is ignored in the following evaluation even though it may not be sufficiently suppressed by the first-generation TDI.

\subsection{Mission sensitivity}

\begin{figure}[htb]
\includegraphics[width=0.46\textwidth]{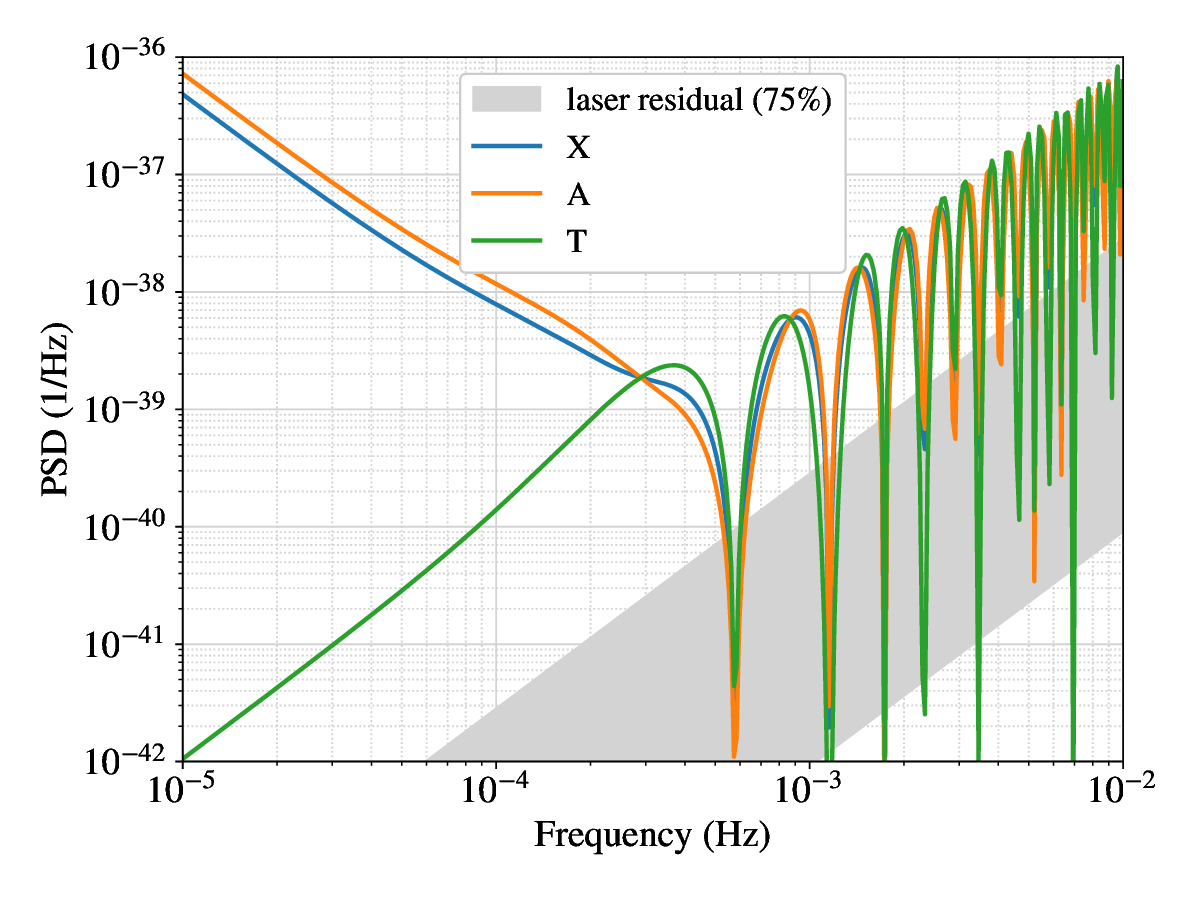} 
\includegraphics[width=0.46\textwidth]{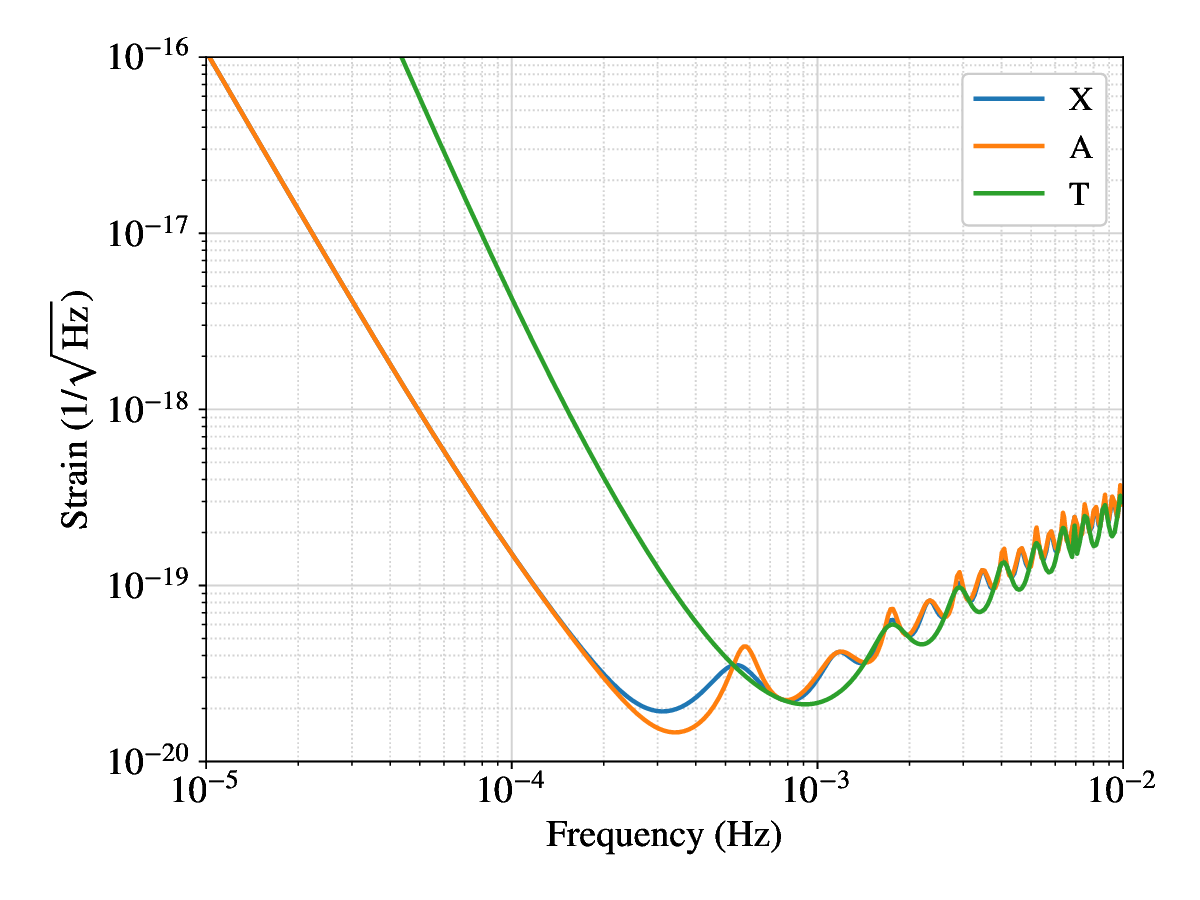} 
\caption{The noise PSDs of the first-generation TDI Michelson channels X, A and T (upper panel), and the average sensitivities of these channels (lower panel) for eASTROD-GW. The curves of E are identical to the A channel and are not shown in the plots. \label{fig:sensitivity}
}
\end{figure}

Excluding the laser noise, the test-mass acceleration noise and the optical metrology noise will be the dominating noises in the TDI channel. The upper limits of acceleration noise and optical metrology noise for the LISA mission are targeted to be \cite{LISA2017},
\begin{align}
 S^{1/2}_{\rm acc, LISA} & = 3 \frac{\rm fm/s^2}{\rm \sqrt{Hz}} \sqrt{ 1 + \left(\frac{0.4 {\rm mHz}}{f} \right)^2 }  \sqrt{ 1 + \left(\frac{f}{8 {\rm mHz}} \right)^4 } , \\
 S^{1/2}_{\rm op, LISA} & = 10 \frac{\rm pm}{ \rm \sqrt{Hz} } \sqrt{ 1 + \left(\frac{2 {\rm mHz}}{f} \right)^4 }.
\end{align}
The noise budgets for elementary ASTROD-GW (eASTROD-GW) are as specified in \cite{Ni:2012eh,Ni:2016wcv} and could be conservatively derived from LISA requirements with different reddening frequencies. The amplitude of acceleration noise is assumed to be equal to the LISA's. The amplitude spectral density (ASD) of optical metrology noise will be 104 times larger than LISA's because of a 104 times larger arm length. Therefore, the noise ASD upper limits for eASTROD-GW are expected to be
\begin{align}
 S^{1/2}_{\rm acc} &= 3 \frac{\rm fm/s^2}{ \rm \sqrt{Hz} } \sqrt{ 1 + \left( \frac{0.1 \ {\rm mHz}}{f} \right)^2 }, \label{eq:b_Sn_acc} \\
 S^{1/2}_{\rm op} & = 1040 \frac{\rm pm}{\rm \sqrt{Hz} } \sqrt{ 1 + \left(\frac{0.2 \ {\rm mHz}}{f} \right)^4 }. \label{eq:b_Sn_op}
 \end{align}
An optimistic configuration for advanced ASTROD-GW (aASTROD-GW) is that the ASDs of the instrumental noises are lower down by one order compared to eASTROD-GW, and the noise budgets could be
\begin{align}
 S^{1/2}_{\rm acc} &= 0.3 \frac{\rm fm/s^2}{ \rm \sqrt{Hz} } \sqrt{ 1 + \left( \frac{0.1 \ {\rm mHz}}{f} \right)^2 }, \label{eq:a_Sn_acc} \\
 S^{1/2}_{\rm op} & = 104 \frac{\rm pm}{\rm \sqrt{Hz} } \sqrt{ 1 + \left(\frac{0.2 \ {\rm mHz}}{f} \right)^4 }. \label{eq:a_Sn_op}
 \end{align}
With the instrumental noises, the noise PSDs of the TDI channels for Michelson X, A, E, and T could be evaluated \cite{1999ApJ...527..814A,2000PhRvD..62d2002E,2001CQGra..18.4059A,Vallisneri:2007xa,Vallisneri:2012np,Wang:1stTDI},
\begin{align}
 S_{\rm n, X}  = & 16 S_{\rm op} \sin^2 x \notag \\ &+ 16 S_{\rm acc} (3 + \cos 2x ) \sin^2 x , \\
 S_{\rm n, A}  = S_{\rm n, E} = & 8  S_{\rm op}  (2+\cos x ) \sin^2 x \notag \\ & + 16  S_{\rm acc}  (3+ 2 \cos x + \cos 2x )  \sin^2 x, \\
 S_{\rm n, T} = & 16  S_{\rm op} (1 - \cos x ) \sin^2 x \notag \\ & + 128  S_{\rm acc} \sin^2 x \sin^4(x/2).
\end{align}
The noise PSDs of three TDI channels for eASTROD-GW are shown in the upper plot of Fig. \ref{fig:sensitivity}. The curve of E is identical to the A and does not show in the plots. The PSD of T channel is significantly lower than X/A/E for frequencies lower than 0.3 mHz, and it becomes comparable to other channels in the higher frequency band. 

On the other side, the GW response of a TDI channel will also depend on the combination of the links. As the curves illustrated in Fig. \ref{fig:response}, the response functions are suppressed by TDI in the lower frequency band for all channels, the T channel is much lower than other channels. And their performances become comparable for frequencies higher than 0.4 mHz. The average sensitivity of a TDI channel is achieved by weighting the noise PSD with its GW response, $S_{\rm TDI} = \sqrt{S_{\rm n, TDI} / \mathcal{R}_{\rm TDI} }$, and the sensitivities of selected channels are shown in the lower panel of Fig. \ref{fig:sensitivity}. 
As we can read from the plot, the sensitivity of A is slightly better than X in the frequencies around 0.4 mHz, and the sensitivities of A and X are consistent in other frequencies. The sensitivity of T observable is much worse than A or X for frequencies lower than 0.5 mHz, and it becomes comparable to the science channels in the higher frequency band. The noise PSDs of the aASTROD-GW will be two orders lower than the eASTROD-GW, and the sensitivity (corresponding to ASD) of aASTROD-GW is one order better than eASTROD-GW's.

\section{Galactic binaries identification} \label{sec:galactic_binaries}

\subsection{galactic binary populations}

The galactic DWD populations used in this study are obtained from \citet{Thiele:2021yyb,thiele_data}. The evolution of binaries populations was simulated using the COSMIC population synthesis suite  \footnote{https://cosmic-popsynth.github.io}, which employs single and binary star evolution algorithms SSE/BSE \cite{Hurley:2000pk,Hurley:2002rf}.
The DWDs were generated based on the assumptions of a fixed binary fraction of 50\% in the populations. The study included four types of DWD which are double helium WDs, a carbon-oxygen WD with a helium WD, double carbon-oxygen WDs, and an oxygen-neon with a companion WD of helium, carbon-oxygen, or oxygen-neon.

\begin{figure}[htb]
\includegraphics[width=0.48\textwidth]{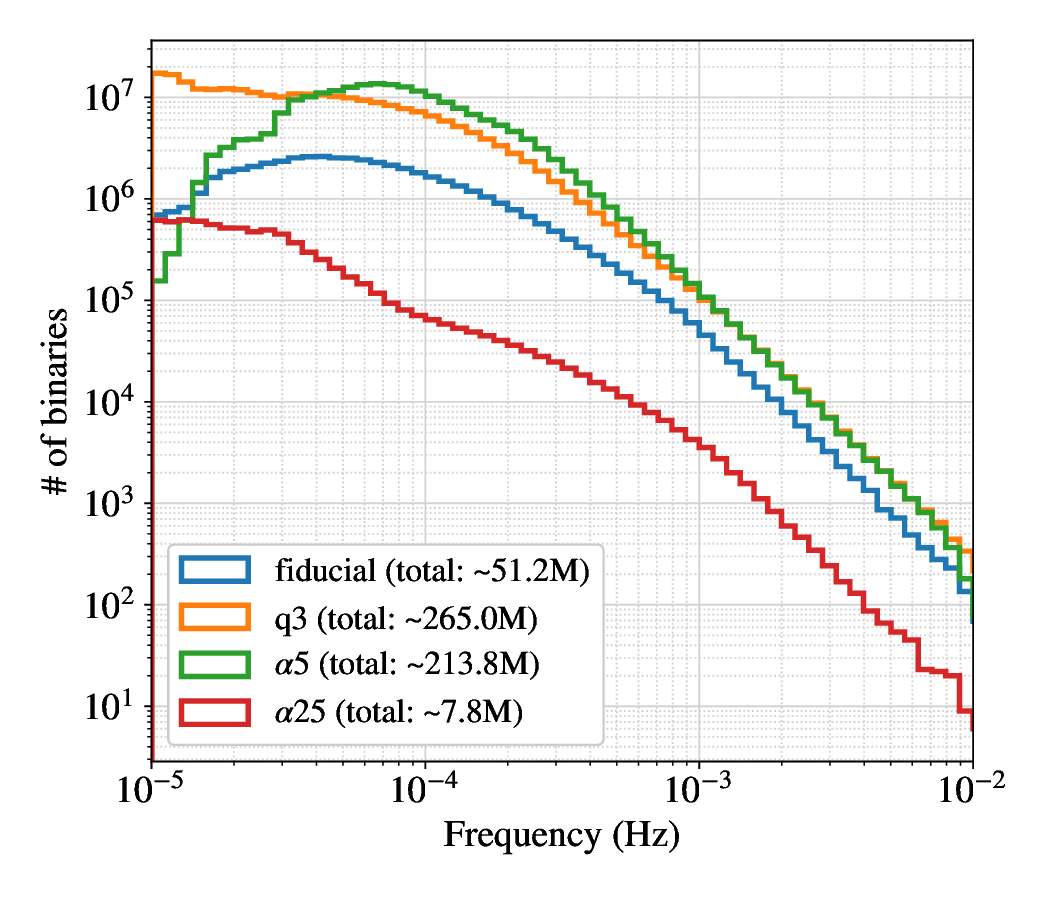} 
\caption{The histograms of the four DWD populations in targeting frequency band. The total number of each population higher than 10 $\mu$Hz is shown in the corresponding legends. \label{fig:pop_hist}
}
\end{figure}

Since the DWD populations vary with the binary evolution assumptions, parameters were utilized to characterize the models of binary evolution. Two variables were employed during the simulations to qualify the binary evolution key phases. The first one is the critical mass ratio $q_\mathrm{c}$ to determine the mass transfer of Roche-lobe overflow, and the other parameter is the efficiency of ejection $\alpha$ during the common envelope. Four populations were generated based on the different parameter setups. The first one was the fiducial case by choosing the \texttt{COSMIC} default settings in \citep{Breivik:2019lmt} with $q_\mathrm{c}=1.6$ and $\alpha=1$. Additional three variant DWD populations were generated by altering these two parameters: in variation $q3$, the factor of $q_\mathrm{c}$ was altered to 3; the $\alpha$5 and $\alpha$25 indicate the populations yielded from the efficiency factor of $\alpha=5$ and $\alpha =0.25$, respectively.
These variations of binary evolution parameter significantly affect the formation efficiency of the DWD. For instance, smaller common envelope ejection efficiencies led to more DWD progenitor mergers where the envelope ejection fails, which is pronounced in comparison between the $\alpha25$ and $\alpha5$. The larger $q_\mathrm{c}$ increased the chance of survival from the common envelope ejection and then led to fewer mergers. As shown in Fig. \ref{fig:pop_hist}, the total number and GW frequency distribution of DWDs differ among the four populations in frequencies higher than 10 $\mu$Hz.

\subsection{Algorithm for galactic binaries identification} \label{subsec:algorithm}

The galactic binaries are mostly in the early inspiral phase, and the frequency evolution of a binary due to GW radiation could be approximated as \cite{Maggiore:2007ulw}, 
\begin{equation}
\dot{f} = \frac{96}{5} \pi^{8/3} \left( \frac{G \mathcal{M}_c }{c^3} \right)^{5/3} f^{11/3},
\end{equation}
where $\mathcal{M}_c = (m_1 m_2)^{3/5} / (m_1+m_2)^{1/5}$ is the chirp mass, $G$ is the gravitational constant and $c$ is the speed of light.
The evolution in lower frequencies is much lower than the higher frequencies, and GWs emitted by the galactic binaries are approximated as monochromatic source (In realistic scenario, the signal will modulate with the orbital motion of the interferometer, and the modulation effect is illustrated in Appendix \ref{sec:modulation_waveform}). The GW waveforms of two polarizations could be described as
\begin{align}
h_{+} & = A ( 1 + \cos^2 \iota ) \cos \Phi,  \\
h_{\times} & =  2 A  \cos \iota \sin \Phi ,
\end{align}
with the amplitude
\begin{equation}
A = \frac{2}{ d_L } \left( \frac{ G \mathcal{M}_c }{ c^2 } \right)^{5/3} \left( \frac{ \pi f }{c} \right)^{2/3},
\end{equation}
where $\iota$ is the inclination of a binary, $\Phi$ is the phase of the GW waveform, $d_L$ is the luminosity distance. 
The power of a monochromatic source in the observation time $T_\mathrm{obs}$ will be,
\begin{equation}
S_h = \frac{ h^2_{+} + h^2_{\times} }{2} T_\mathrm{obs}.
\end{equation}
And the corresponding signal-to-noise ratio (SNR), $\rho$, with the confusion noise could be obtained by implementing,
\begin{equation} \label{eq:SNR}
\rho^2 =  \sum_\mathrm{A, E, T}  \frac{2 S_h}{ S_\mathrm{inst, TDI} + S_\mathrm{conf} },
\end{equation}
where $S_\mathrm{inst, TDI} $ is the instrumental noise PSD of the TDI channel, and $S_\mathrm{conf} $ is the PSD of foreground noise.

To identify binaries from a population, an algorithm derived from \cite{Karnesis:2021tsh} is implemented. Since the sub-mHz mission(s) would be scheduled after the LISA, a hierarchical scenario is considered. For a binary population, the binary identifications by employing LISA are performed at first, and the resolvable sources from LISA's 6-year observation are removed from the population. The unresolvable binaries remain as the initial population for ASTROD-GW, the binaries resolved by the ASTROD-GW in 10-year observation are further removed from the population. Then the indistinguishable binaries are utilized to evaluate the confusion noise for the ASTROD-GW mission. The specific steps for the algorithm are as follows.
\begin{itemize}
\item[a)] After setting the mission (LISA or ASTROD-GW) sensitivity and observation duration, the frequency bins are generated based on the frequency resolution, $f_i = [ i \Delta f, (i+1) \Delta f ), \ (i = 0, 1, 2... n) $, where $\Delta f = 1 / T_\mathrm{obs}$. Each binary is classified based on the frequency bins they belong to. 
\item[b)] The PSD of GWs from binaries in each frequency bin is calculated at first, then a confusion noise curve is smoothed by averaging spectra over adjacent frequency bins. And a preliminary noise function of frequency is constructed by interpolating their mean values.
\item[c)] The SNR is calculated for each binary by using Eq. \eqref{eq:SNR}, and the sources with $\rho \geqslant 7$ are labeled as resolved and removed from the population. The confusion noise function is updated from the left unresolved binaries.
The SNRs of previously unresolved binaries are recalculated by utilizing the updated observational noise function, $S_\mathrm{n, obs} =  S_\mathrm{inst} + S_\mathrm{conf}$.
The additional binaries with $\rho \geqslant 7$ are labeled as resolvable and subtracted, and the confusion noise spectrum is reevaluated. 
The iteration is run until no additional binary becomes resolvable with the latest confusion noise.
\item[d)] After the evaluation is performed for the LISA mission, the unresolved sources are initialized as the aimed population for the ASTROD-GW, and the algorithm is rerun for a sensitivity configuration of eASTROD-GW or aASTROD-GW.
\end{itemize}

\subsection{Evaluations of galactic confusion noise}

\begin{figure*}[htb]
\includegraphics[width=0.42\textwidth]{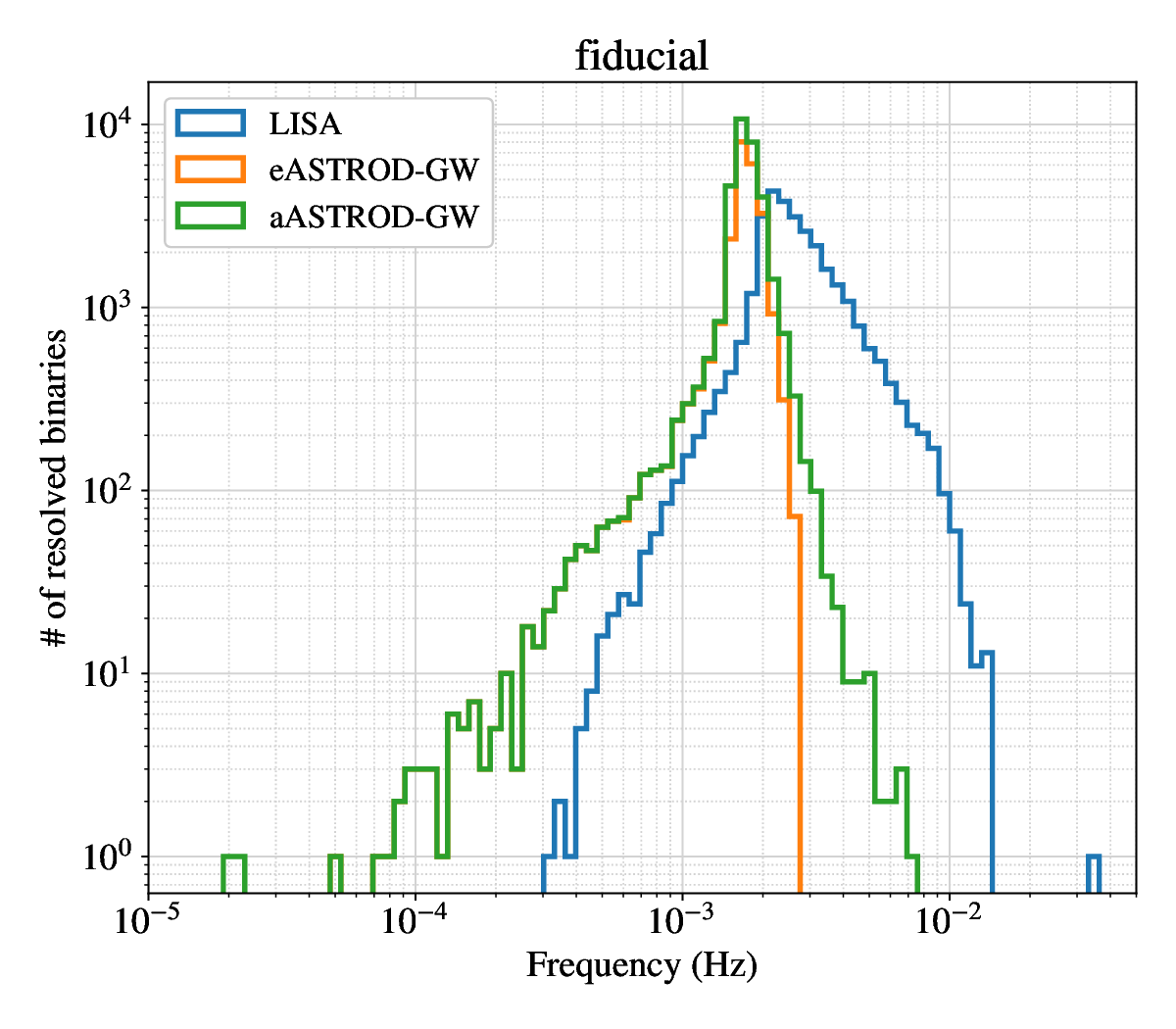} 
\includegraphics[width=0.42\textwidth]{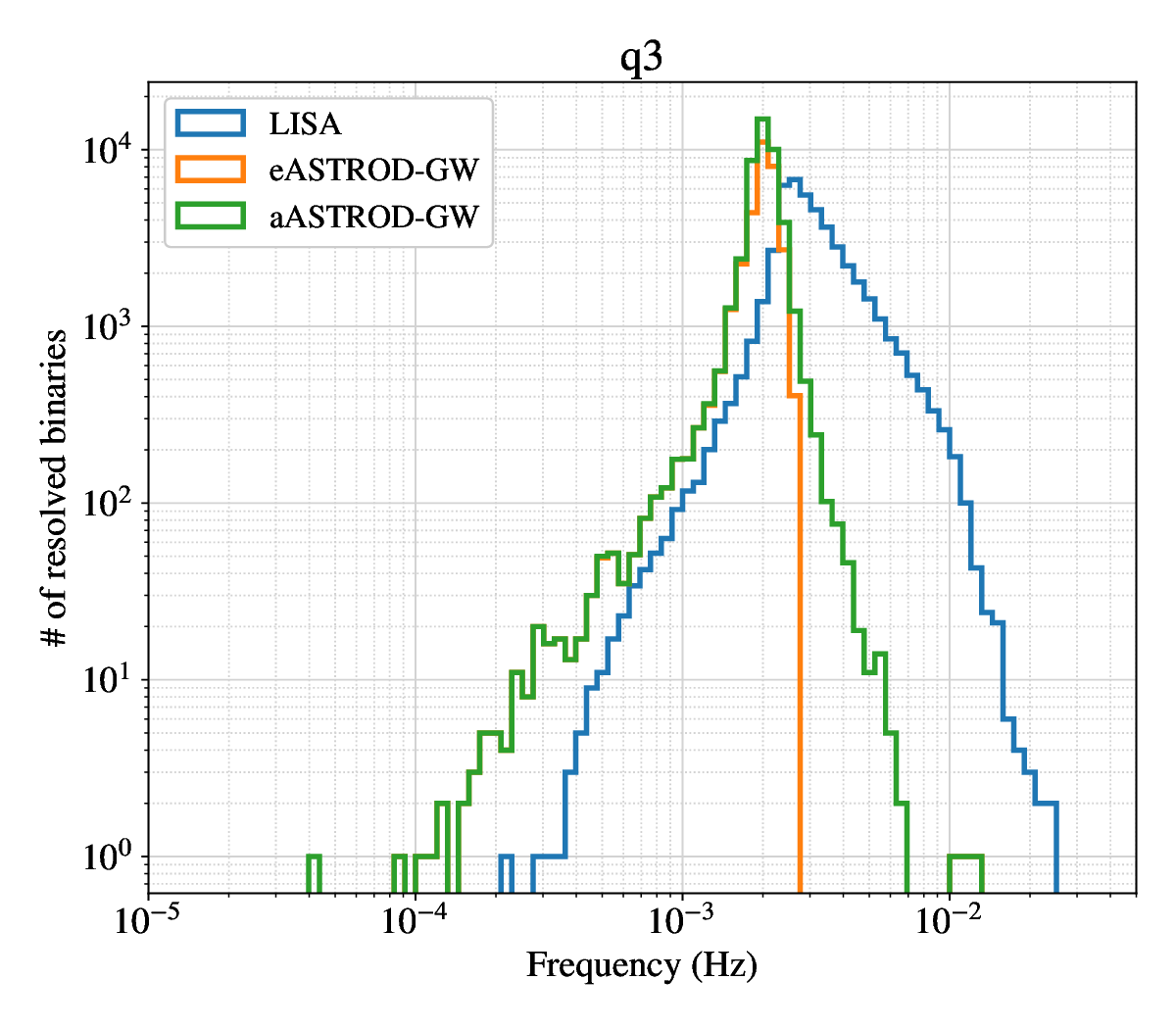} 
\includegraphics[width=0.42\textwidth]{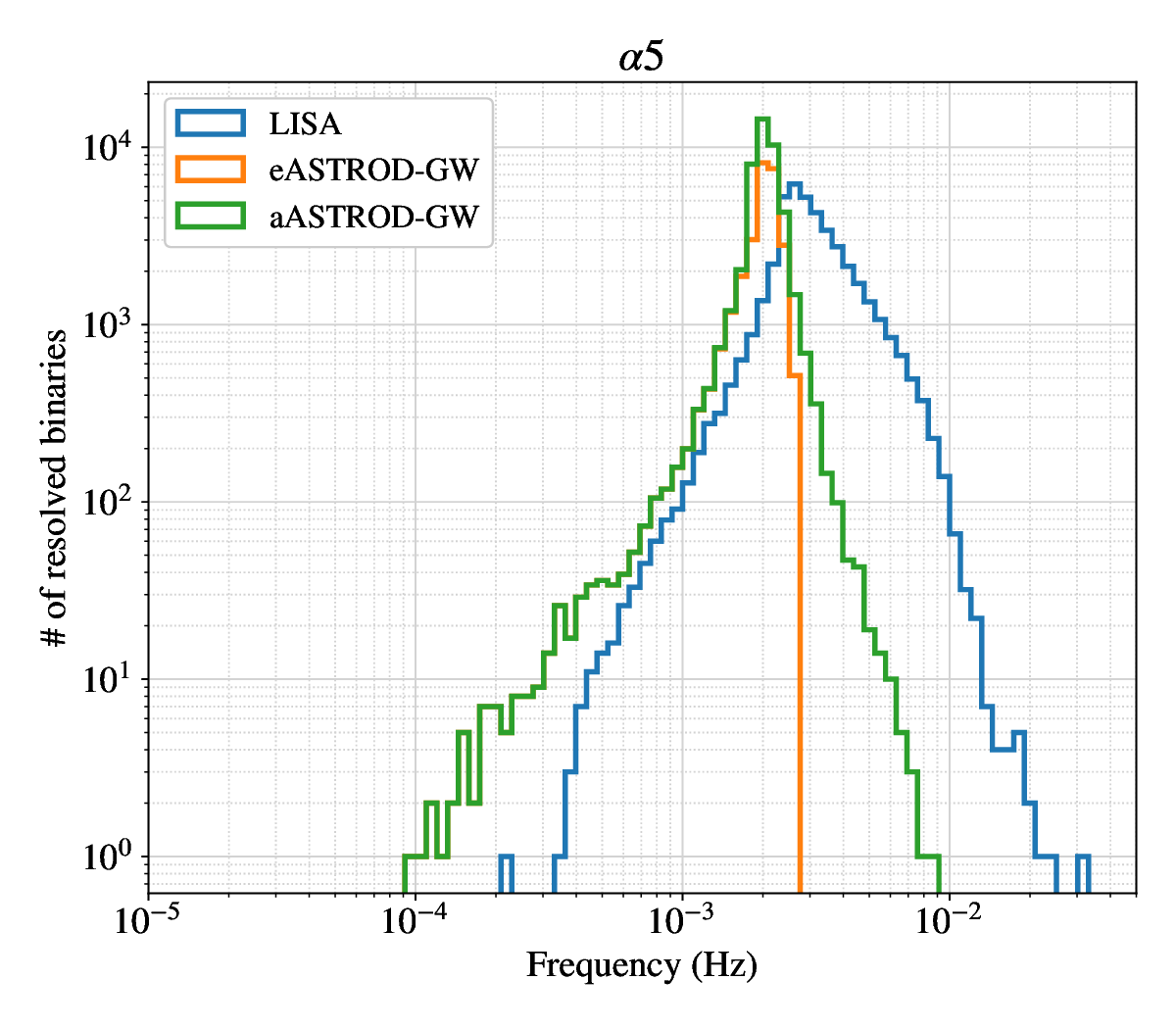} 
\includegraphics[width=0.42\textwidth]{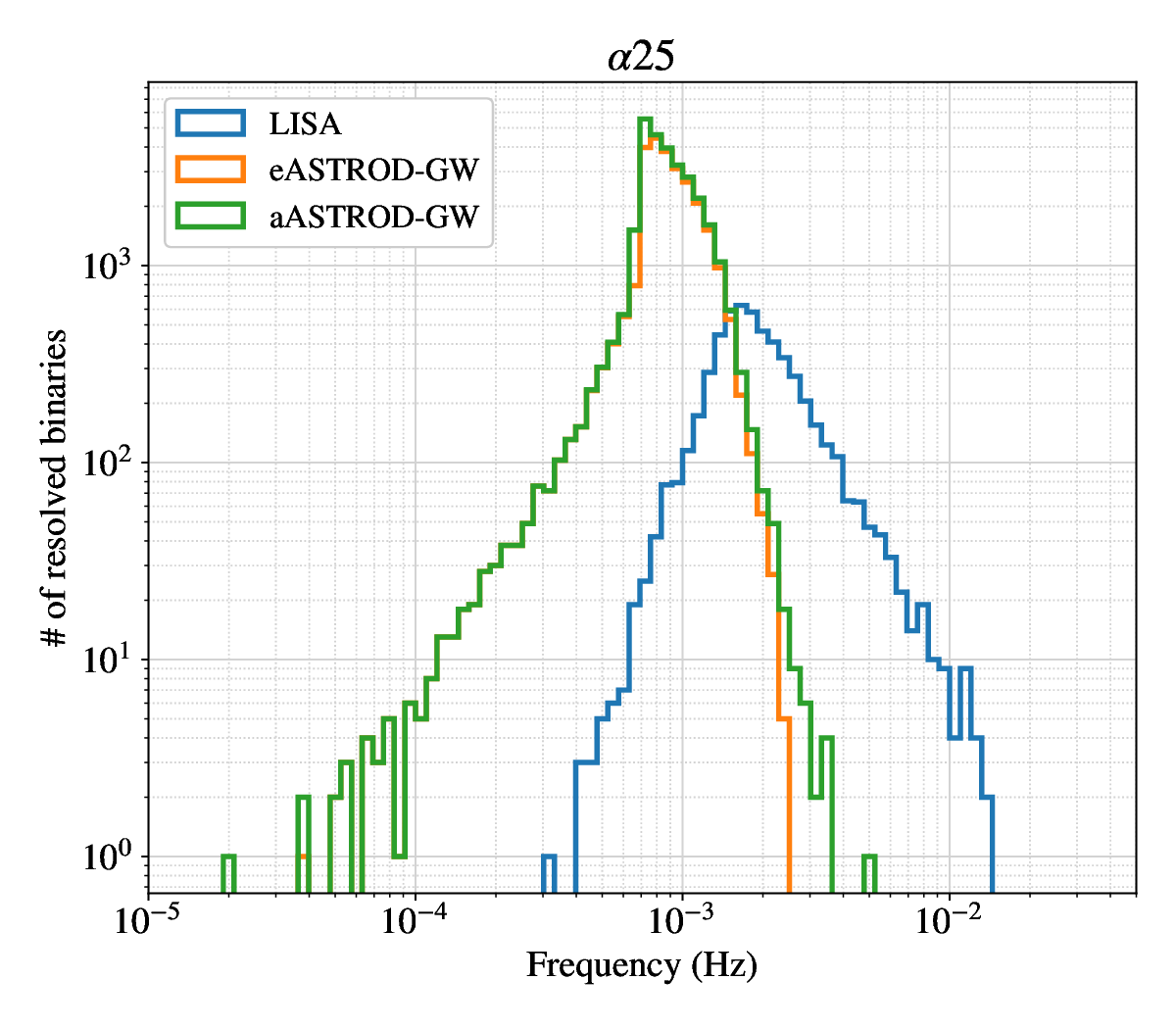}
\caption{The histograms of the resolved DWDs for LISA in 6 years of observation (blue) and eASTROD-GW (orange) or aASTROD-GW (green) in 10 years of observation. The resolved binaries by LISA are mostly at higher frequencies because they are around LISA's sensitive range and (almost) free of confusion noise. Beyond the resolved sources from LISA, the ASTROD-GW could further identify the binaries in a lower frequency range. The more sensitive aASTROD-GW could detect fainter DWDs in a range of $\sim$[1, 10] mHz compared to eASTROD-GW. \label{fig:hist_resolved} }
\end{figure*}

\begin{figure*}[htb]
\includegraphics[width=0.49\textwidth]{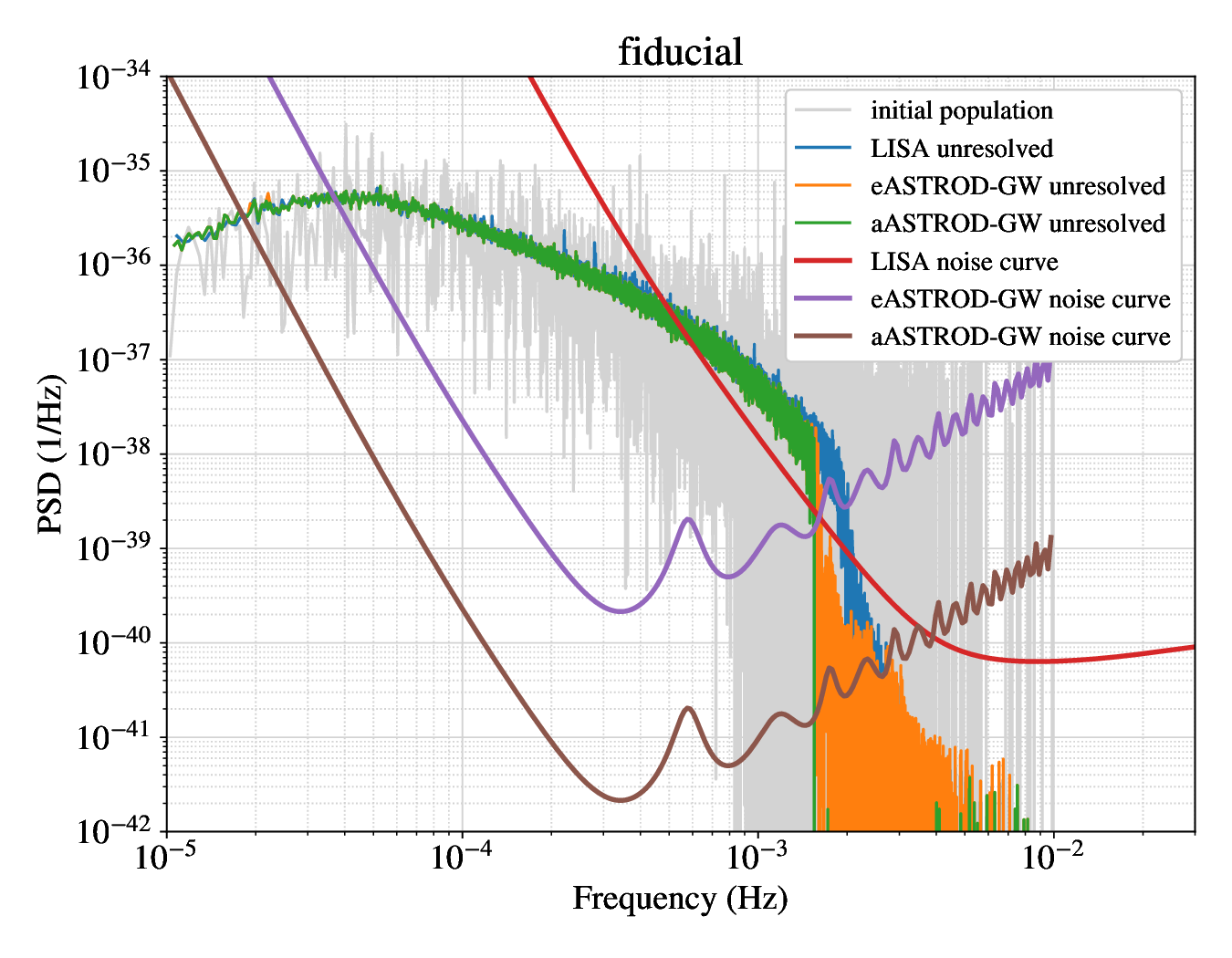} 
\includegraphics[width=0.49\textwidth]{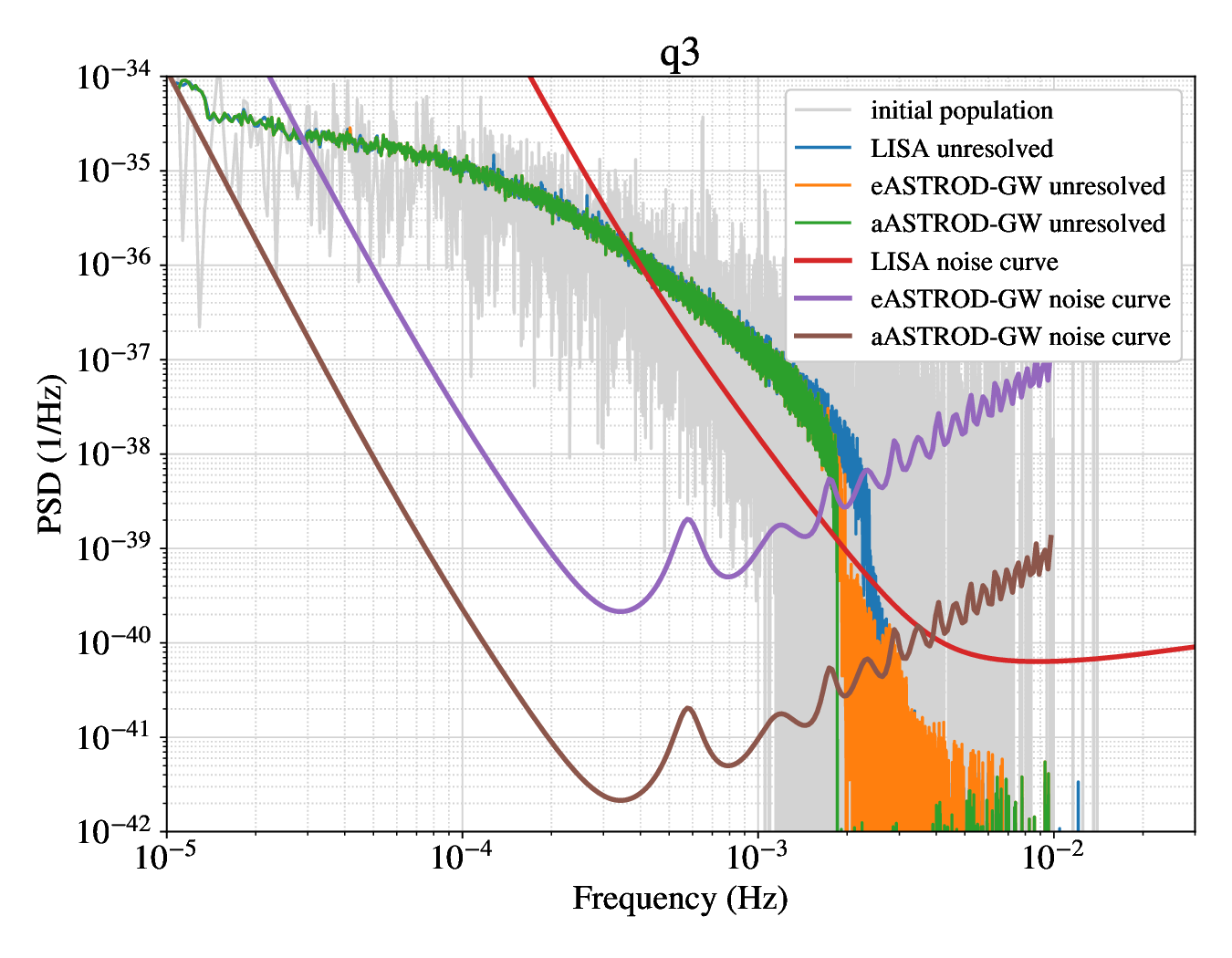} 
\includegraphics[width=0.49\textwidth]{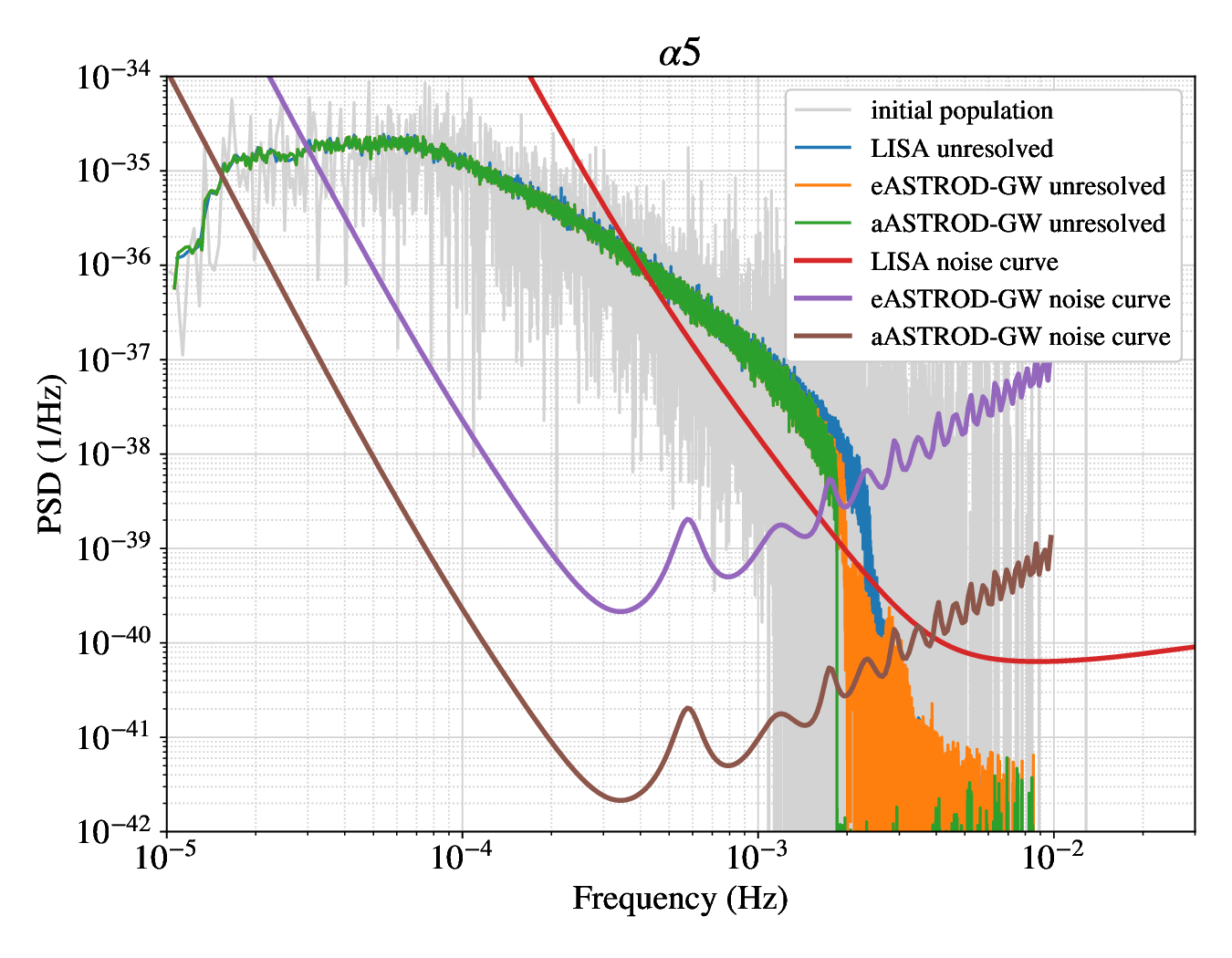} 
\includegraphics[width=0.49\textwidth]{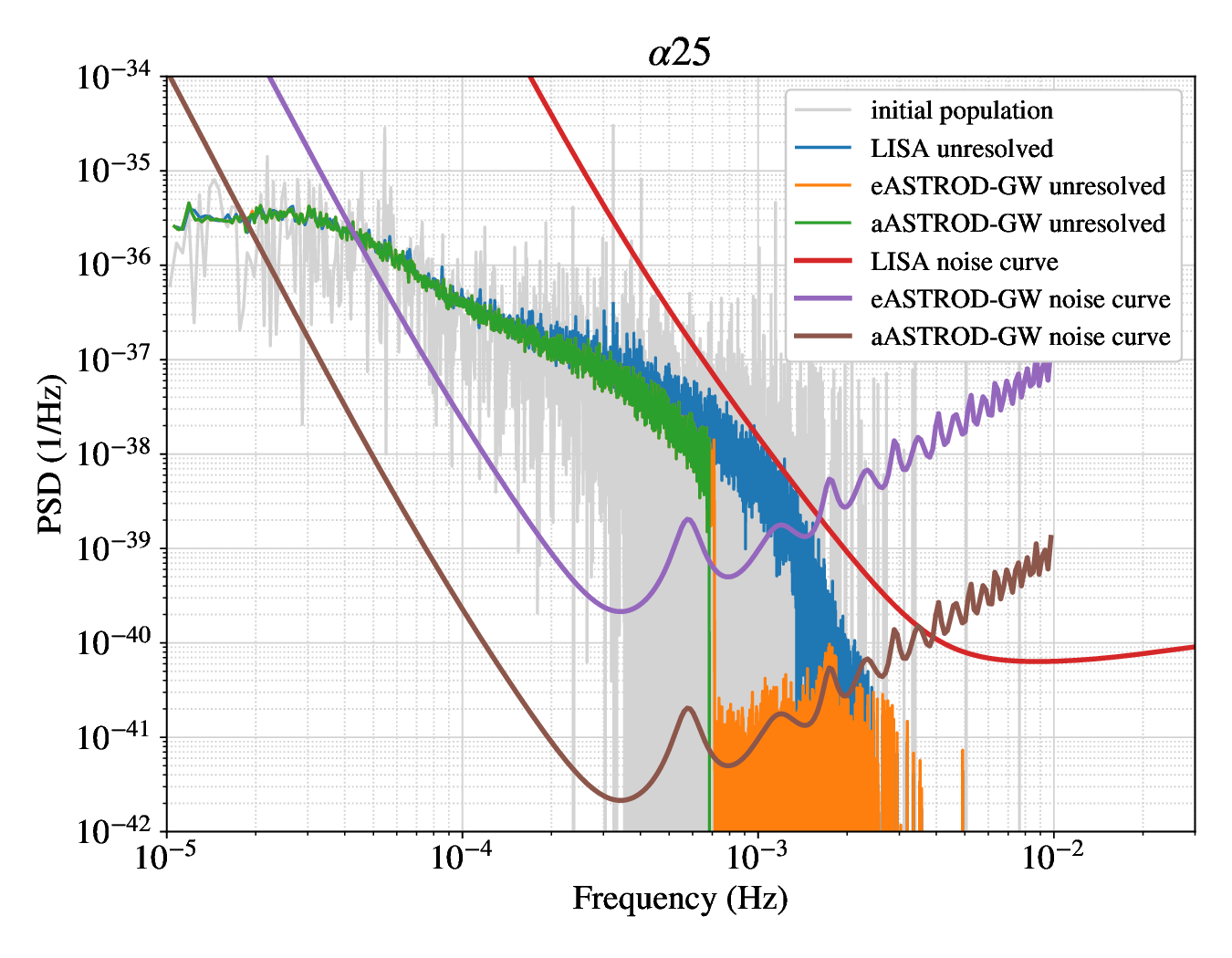}
\caption{The initial DWD population (grey), the unresolved foreground for LISA in 6 years of observation (blue), and the unresolved foreground for eASTROD-GW (orange) or aASTROD-GW (green) in 10 years of observation. The sensitivities of the first-generation TDI Michelson-A for LISA, eASTROD-GW, and aASTROD-GW are shown by red, purple, and brown curves, respectively. \label{fig:pop_foreground_ASTROD} }
\end{figure*}

The numbers of the DWD populations and hierarchically resolved binaries by LISA and ASTROD-GW are shown in Table \ref{tab:N_resolved}, and the histograms of the resolved sources are shown in Fig. \ref{fig:hist_resolved}.
For the LISA, the resolved binaries are mostly at higher frequencies. One reason is that these frequencies are the sensitive band of the LISA, and another reason would be that the binaries in the higher frequency band are sparsely distributed which will be barely affected by confusion noise. In the four populations, there are more DWDs in the higher frequency range for the larger population as illustrated in Fig. \ref{fig:pop_hist}. As a result, more than $\sim$46,000 of the binaries could be identified by the LISA for the largest q3 population, the number of resolvable binaries in $\alpha$5 and fiducial will be $\sim$43,000 and $\sim$30,000, respectively. Only $\sim$5,500 binaries would be detected for the least population $\alpha$25. After the identified binaries are subtracted from the populations, the power spectra formed by unresolved binaries for LISA are shown by blue curves in Fig. \ref{fig:pop_foreground_ASTROD}. The most significant confusion noises are yielded by the q3 and $\alpha$5 cases, and the confusion noise from $\alpha$25 should be trivial for LISA's observation.

\begin{table}[tbh]
\caption{\label{tab:N_resolved} The numbers of the DWD populations for LISA and ASTROD-GW, and the numbers of revolved binaries by the LISA in 6-year observation and by the ASTROD-GW in 10-year observation.
}
\begin{ruledtabular}
\begin{tabular}{ccccc}
 &  fiducial & q3 & $\alpha$5 &  $\alpha$25  \\
\hline
\# of population & 51,216,193 & 265,024,921 & 213,775,451 & 7,803,813 \\
LISA & 30244 & 46527 & 43086 & 5506 \\
\hline
\# of population & 51,185,949 & 264,978,394 & 213,732,365 & 7,798,307 \\ 
eASTROD-GW & 24240 & 32296 & 27608 & 26595 \\
aASTROD-GW & 33388 & 45609 & 45733 & 30060 \\
\end{tabular}
\end{ruledtabular}
\end{table}

With the loud and dispersed DWDs identified by the LISA and removed from populations, the populations for ASTROD-GW are more concentrated in the lower frequency range, and the ASTROD-GW could further detect extra binaries in lower frequencies. Besides the aforementioned relatively sparser sources around 1 mHz, the higher SNR and finer frequency resolution achieved from the longer observation duration should be counted, as well as the sensitive frequencies of the detector.
One typical case is from the $\alpha$25 population, compared to the $\sim$5,500 binaries detected by LISA, the ASTROD-GW could identify $\sim$26,000 and $\sim$30,000 in elementary and advanced configurations, respectively. This is because more binaries are sparsely distributed in the frequencies lower than $\sim$1 mHz which is approaching the ASTROD-GW's sensitive band as shown in the lower right panel of Fig. \ref{fig:pop_foreground_ASTROD}. For the other three populations, $\sim$24,000 to $\sim$32,000 binaries are resolved by eASTROD-GW which relate to the numbers of populations around 2 mHz. With better sensitivity, the aASTROD-GW could identify more fainter sources in the frequency range of $\sim$[1, 10] mHz than the eASTROD-GW as shown in Fig. \ref{fig:hist_resolved} and Fig. \ref{fig:pop_foreground_ASTROD}.
The aASTROD-GW configuration does not promote the numbers of detection in the frequencies lower than $\sim$1 mHz compared to the eASTROD-GW, because these sub-mHz missions substantially are subject to the galactic confusion noise in this frequency band. 

The resolved DWDs are a small fraction of the population, and most of the binaries are unsolvable. The GWs from these sources will form a significant confusion noise for the sub-mHz observation, and the PSD of confusion noise could be orders higher than the instrument noises. The ASTROD-GW for either elementary or advanced configuration will essentially face the confusion noise in the same frequency ranges at a same levels as shown in \ref{fig:pop_foreground_ASTROD}. 
The GW spectra of unresolved binaries could be fitted with a polynomial formation in a log-log scale \cite{Thiele:2021yyb},
\begin{equation} \label{eq:fitting_model}
\log_{10} S_\mathrm{conf} = \sum^{4}_{k=0} a_k \left( \log_{10} f \right)^k.
\end{equation}
The confusion noise spectra for ASTROD-GW are fitted with frequency cutoffs of 1.6 mHz for fiducial, 1.7 mH for q3 and $\alpha$5, and 0.6 mHz for $\alpha$25. And five coefficients of $a_k$ for four population models are shown in Table \ref{tab:fitted_paras}. 

\begin{table}[tbh]
\caption{\label{tab:fitted_paras} The fitted coefficients of the galactic foreground spectra for four populations.
}
\begin{ruledtabular}
\begin{tabular}{ccccc}
 parameter &  fiducial & q3 & $\alpha$5 &  $\alpha$25  \\
\hline
 $a_0$ & -180.460 & -115.440 & -170.709 & -447.183 \\
 $a_1$ & -145.710 & -72.8789 & -137.796 & -414.383 \\
 $a_2$ & -56.2753 & -25.8721 & -54.3750 & -157.453 \\
 $a_3$ & -9.80524 & -4.22458 & -9.75381 & -26.5968 \\
 $a_4$ & -0.64482 & -0.26361 & -0.66272 & -1.67843 \\
\end{tabular}
\end{ruledtabular}
\end{table}

\section{Galactic foreground characterization} \label{sec:characterization}

\subsection{Algorithm for characterization}

The presence of confusion noise in the GW observation data would be like a stochastic process. From another perspective, the confusion noise would be a dominating stochastic GW foreground signal in the sub-mHz band.
For the LISA mission, the normal direction of the spacecraft constellation will change will time, and its antenna pattern to the galactic foreground will have yearly modulation. For mission orbits like ASTROD-GW and Folkner missions, their directions of the constellation will be (nearly) constant in the mission duration, and the response to these sources would not change as significantly as LISA. Although the spatial distributions of the galactic binaries are anisotropic, the appearance of the foreground in the observation data may not be strongly modulated for these sub-mHz detectors. The determination of the foreground is could be examined by assuming it is stationary. With the fitted models of foreground spectra, we perform the foreground inference by using the simulated observation data. 

The instrumental noises (acceleration noise and optical metrology noise) are assumed to be Gaussian and stationary, and they are generated for each optical bench in the time domain based on the noise budgets in Eqs. \eqref{eq:b_Sn_acc}-\eqref{eq:a_Sn_op}. And the single link measurements are combined by using Eqs. \eqref{eq:s_epsilon_tau_1}-\eqref{eq:eta}. After that, TDI data of Michelson (X, Y, Z) are synthesized with the time-shifted single links measurements, and the optimal channels are obtained by implementing Eq. \eqref{eq:optimalTDI}. Three optimal TDI channels, (A, E and T), are employed to perform the galactic foreground characterization in our algorithm. In principle, the duty cycle of a space mission would be 75\%, and the observation data would have gaps. Considering the gap would not be significant for the stochastic GW analysis, the data is generated consecutively in 10 years.

The sampling frequency is set to be 20 mHz which corresponds to a Nyquist frequency of 10 mHz. For the foreground, the affected frequencies are mostly lower than $\sim$2 mHz. To reduce the computing time, the high-frequency cutoffs are selected to be at the frequency cutoffs of the foreground.
For the eASTROD-GW configuration, the high-frequency limit for the $\alpha$25 model is set to be 0.7 mHz, the high boundary for the fiducial model is 1.6 mHz, and the cutoffs for the $q3$ and $\alpha$5 are set to be 1.8 mHz. The cutoffs for the aASTROD-GW should be slightly lower than the eASTROD-GW's as could be distinguished from Fig. \ref{fig:pop_foreground_ASTROD}, and the frequencies are also tuned in the algorithm. 

With the mock data, five parameters of the foreground are estimated by utilizing the Bayesian algorithm, and the instrument noises are assumed to be known. As discussed in Sec. \ref{subsec:orbit}, the arm lengths of the mission are close to equal, then the correlation noise between the TDI channels A, E, and T are ignored. The likelihood function of parameter inference will be \cite{Adams:2010vc,Adams:2013qma}
\begin{equation}
\ln \mathcal{L} ( \vec{ a } ) \propto - \frac{1}{2} \sum_i \left[  \tilde{ \mathbf{s} }^{\dagger} (f_i) \mathbf{\Sigma}^{- 1} (f_i)  \tilde{ \mathbf{s} } (f_i) + \ln | \mathbf{\Sigma} | \right] ,
\end{equation}
where $ \tilde{ \mathbf{s} }$ is the data vector of TDI channels, $\Sigma$ is correlation matrix, 
\begin{equation}
\mathbf{\Sigma} = 
\begin{bmatrix}
S_\mathrm{n,A} + \mathcal{R}_\mathrm{A} S_\mathrm{conf} & 0 & 0 \\
0 & S_\mathrm{n,E} + \mathcal{R}_\mathrm{E} S_\mathrm{conf} & 0 \\
0 & 0 & S_\mathrm{n,T} + \mathcal{R}_\mathrm{T} S_\mathrm{conf} \\
\end{bmatrix},
\end{equation}
and $\mathcal{R}_\mathrm{TDI}$ is the average GW response of a TDI channel as described in Appendix \ref{sec:appendix_response}. The posterior probability is proportional to the product of likelihood and prior $\pi ( \vec{ a } )$,
\begin{equation}
p ( \vec{ a } )  \propto \pi( \vec{ a } ) \mathcal{L} ( \vec{ a } ).
\end{equation}
The prior for each parameter is set to be uniform in a selected range. The Markov chain Monte Carlo sampler in \texttt{emcee} is utilized to run the Bayesian inference \cite{emcee}.
One caveat is that there are characteristic frequencies at $f = n/(2L) \simeq 0.577 n$ mHz $(n = 1, 2, 3...)$ as shown in Fig. \ref{fig:sensitivity} and \ref{fig:response}. These null frequencies should be gated during the analysis, otherwise the corrupted correlation matrix may cause the error for the parameter estimation.

\subsection{Characterization of galactic foreground}

\begin{figure}[htb]
\includegraphics[width=0.48\textwidth]{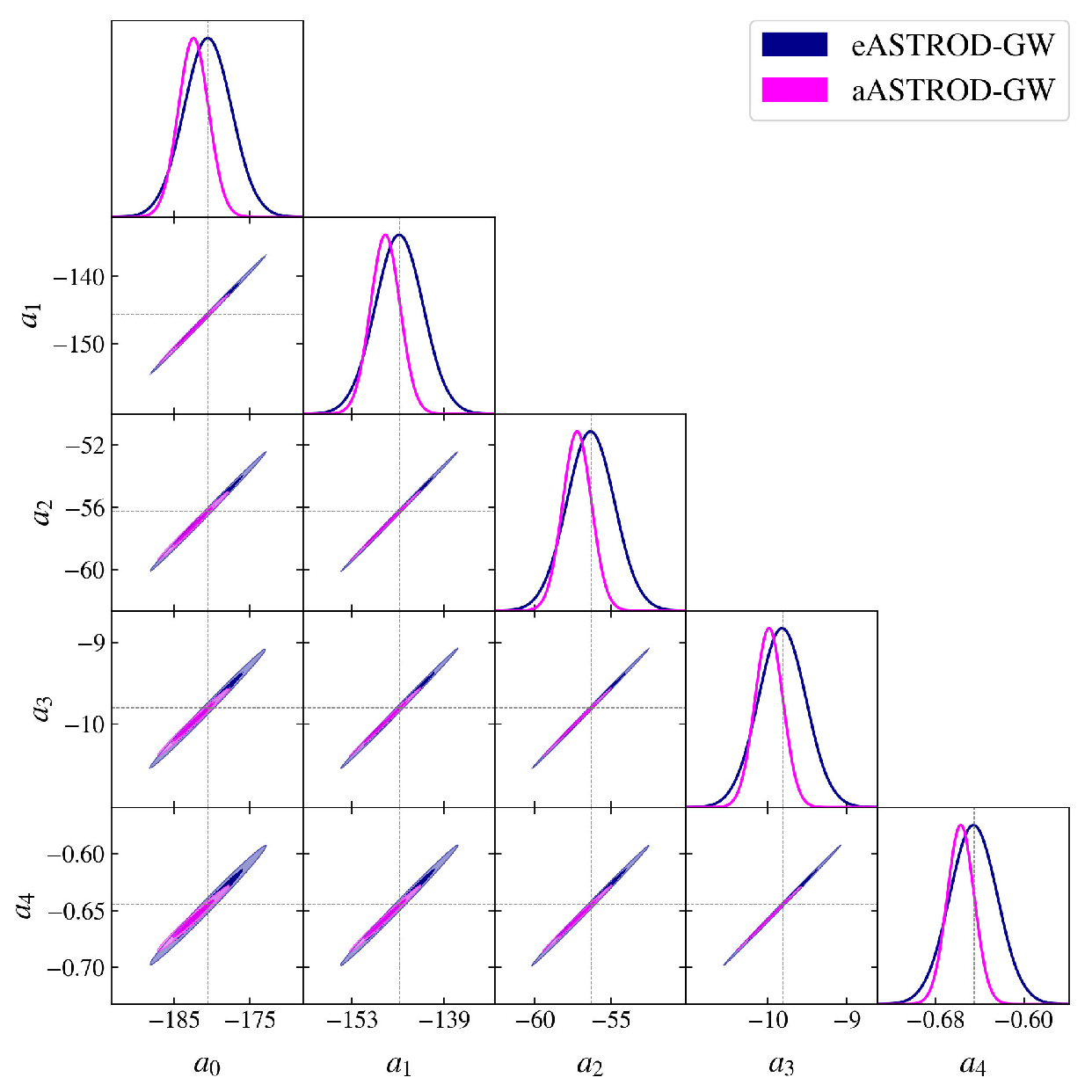}
\caption{\label{fig:corner_fiducial} The corner plot for parameters from the fiducial model inferred from the eASTROD-GW and aASTROD-GW configurations. The uncertainties of parameters from the aASTROD-GW are smaller than eASTROD-GW because of its better sensitivity and wider observable frequency range of the confusion noise as shown in Fig. \ref{fig:pop_foreground_ASTROD}.
}
\end{figure}

The inferred values by using the eASTROD-GW and aASTROD-GW are shown in Table \ref{tab:inferred_paras}. The uncertainties of parameters inferred from the aASTROD-GW case are smaller than the results achieved from eASTROD-GW. The reasons are that the sensitivity of the advanced configuration is better than the elementary case, and that a larger frequency range of foreground is in the sensitive band of aASTROD-GW. 
Comparing the results for different populations, the parameters of the q3 and $\alpha$5 models are better constrained for their strongest foreground, and the parameters of $\alpha$25 are worst inferred corresponding to its lowest foreground level in the four populations. On the other hand, generally speaking, the coefficients of lower orders are better measured than the coefficients of higher orders, for instance, for the results of the fiducial model inferred from the eASTROD-GW case, the relative precision of $a_0$ is $6.2/180 \simeq 0.03$, and the relative uncertainty of $a_4$ is up to $0.043/0.645 \simeq 0.07$.
The results for the fiducial model are selected for comparison as shown in Fig. \ref{fig:corner_fiducial}. As we can see in the plot, besides the parameters inferred from aASTROD-GW are better constrained than the eASTROD-GW, the five parameters are also highly correlated because the parameters degenerate in the log-linear fitting model.

\begin{table*}[tbh]
\caption{\label{tab:inferred_paras} The inferred values of confusion noise parameters by using the eASTROD-GW and aASTROD-GW configurations for four DWD population models. The uncertainty of each parameter is in the range of 1$\sigma$.
}
\begin{ruledtabular}
\begin{tabular}{cccccc}
configuration & parameters &  fiducial & q3 & $\alpha$5 &  $\alpha$25  \\
\hline
& $a_0$ & $-180.478_{-3.130}^{3.127}$ & $-113.321_{-2.146}^{2.143}$ & $-168.240_{-2.306}^{2.303}$ & $-453.998_{-16.820}^{16.638}$ \\
& $a_1$ & $-145.761_{-3.639}^{3.636}$ & $-70.544_{-2.488}^{2.483}$ & $-134.989_{-2.683}^{2.678}$ & $-421.452_{-18.310}^{18.098}$ \\
eASTROD-GW 
& $a_2$ & $-56.309_{-1.577}^{1.576}$ & $-24.917_{-1.074}^{1.073}$ & $-53.187_{-1.162}^{1.158}$ & $-160.196_{-7.450}^{7.355}$ \\
& $a_3$ & $-9.813_{-0.302}^{0.301}$ & $-4.052_{-0.205}^{0.204}$ & $-9.532_{-0.222}^{0.221}$ & $-27.066_{-1.343}^{1.325}$ \\
& $a_4$ & $-0.645_{-0.022}^{0.021}$ & $-0.252_{-0.015}^{0.014}$ & $-0.647_{-0.016}^{0.016}$ & $-1.708_{-0.091}^{0.089}$ \\
\hline
& $a_0$ & $-182.363_{-1.990}^{1.989}$ & $-114.433_{-1.377}^{1.384}$ & $-168.300_{-1.587}^{1.593}$ & $-448.290_{-10.291}^{10.263}$ \\
& $a_1$ & $-147.884_{-2.254}^{2.251}$ & $-71.707_{-1.552}^{1.563}$ & $-135.064_{-1.806}^{1.811}$ & $-415.439_{-10.740}^{10.692}$ \\
aASTROD-GW 
& $a_2$ & $-57.202_{-0.950}^{0.949}$ & $-25.368_{-0.650}^{0.655}$ & $-53.223_{-0.764}^{0.766}$ & $-157.823_{-4.179}^{4.162}$ \\
& $a_3$ & $-9.980_{-0.176}^{0.176}$ & $-4.130_{-0.120}^{0.121}$ & $-9.540_{-0.142}^{0.143}$ & $-26.652_{-0.720}^{0.717}$ \\
& $a_4$ & $-0.657_{-0.012}^{0.012}$ & $-0.257_{-0.008}^{0.008}$ & $-0.648_{-0.010}^{0.010}$ & $-1.681_{-0.046}^{0.046}$ \\
\end{tabular}
\end{ruledtabular}
\end{table*}

As the plots shown in Fig. \ref{fig:pop_foreground_ASTROD}, the galactic foreground overwhelms the instrument noises and other GW signals in observation data for a sub-mHz mission. In an optimistic assumption, if the foreground could be well modeled and characterized, the foreground may be subtracted from the data, and the sensitivity of GW observation may be improved. To estimate the residual after the foreground subtraction, the spectra of the foreground are restored. As the first step, 5000 samples of the foreground parameters, $\alpha_i$, are randomly picked from the MCMC samples, and the galactic foreground is calculated by using Eq. \eqref{eq:fitting_model}; then the distribution of foreground spectrum at each frequency is obtained, and the $1\sigma$ confidence intervals of the spectra are evaluated for each frequency in the sensitive band. We optimistically treated the difference between the central values and the $1\sigma$ boundaries as the residual of the foreground subtraction.

\begin{figure}[htb]
\includegraphics[width=0.48\textwidth]{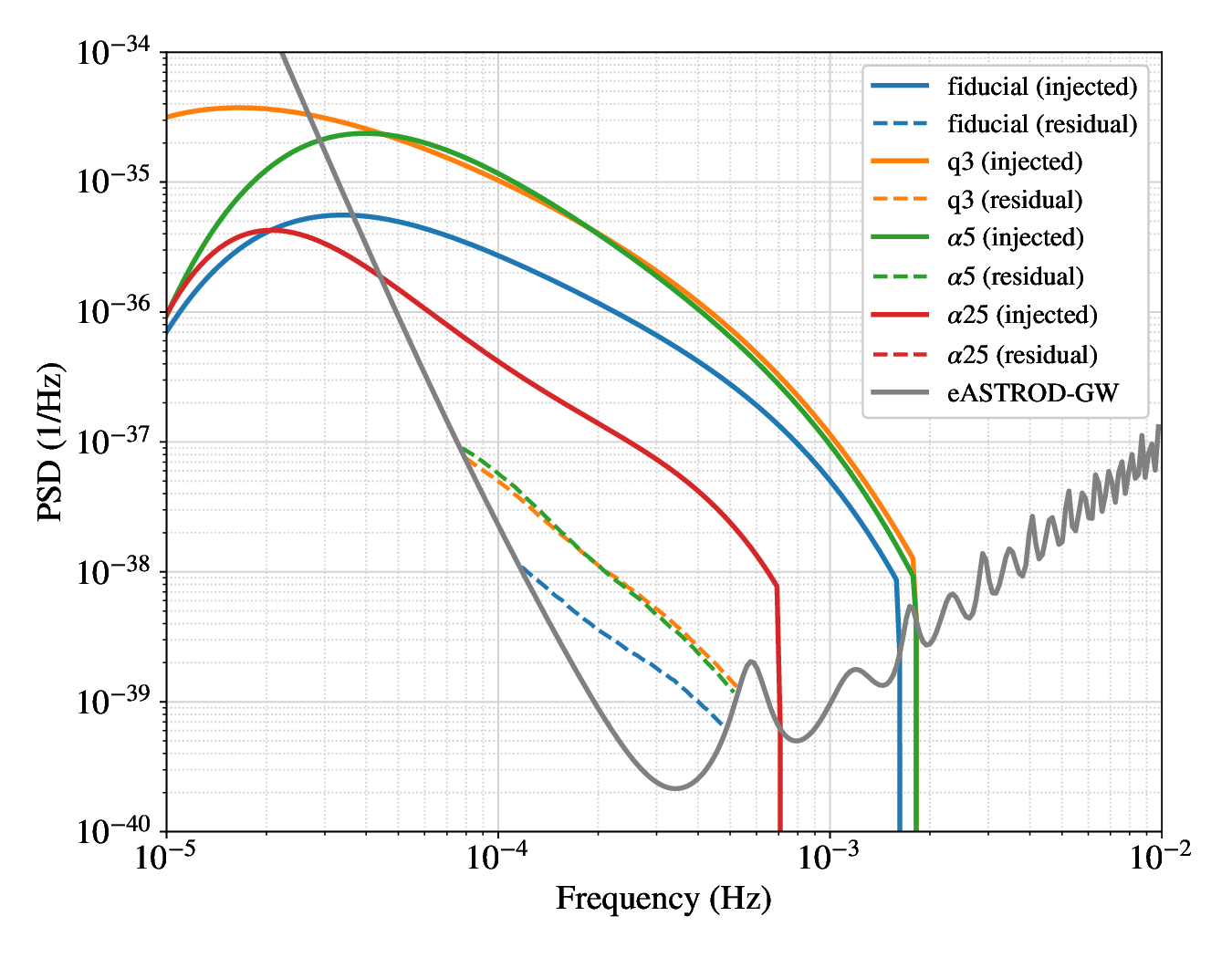}
\includegraphics[width=0.48\textwidth]{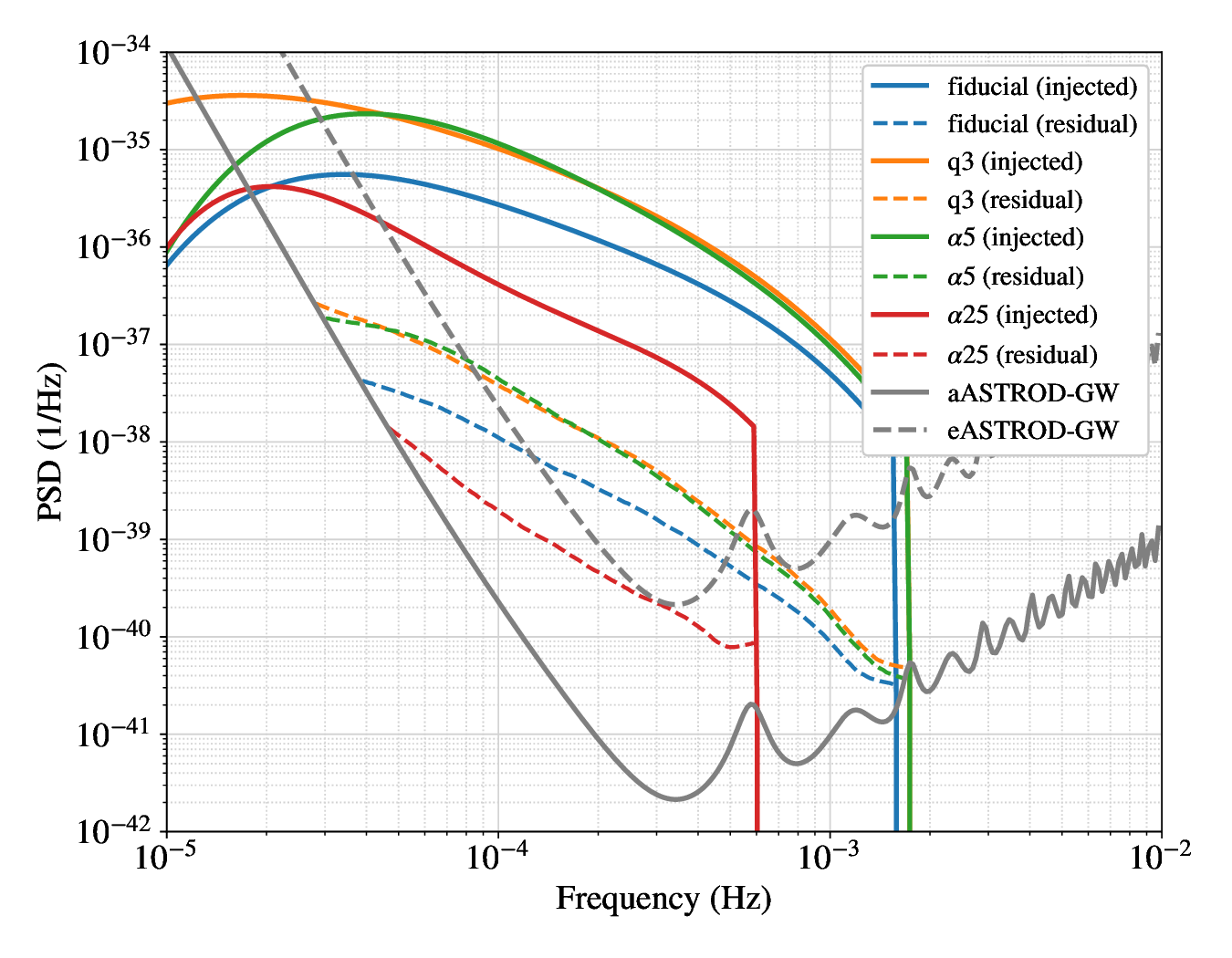}
\caption{\label{fig:ASTROD_GW_fg_res} The injected foregrounds (solid color lines) and corresponding residuals (dashed color lines) for the eASTROD-GW (upper) and aASTROD-GW (lower) configurations. The residuals are obtained from differences between central values of the restored foreground spectra and boundaries of the $1\sigma$ confidence intervals. (in the upper panel, the residual of $\alpha$25 could not be seen because it is beneath the eASTROD-GW noise curve).
}
\end{figure}

The injected foregrounds (solid lines) and the corresponding residuals (dashed lines) for eASTROD-GW and aASTROD-GW are shown in the upper and lower panel of Fig. \ref{fig:ASTROD_GW_fg_res}, respectively. For both cases, the residuals would be more than two orders lower than the injected foreground in the target frequency band. Since we are concerned with the foreground in which frequencies are higher than the instrumental noises, the residual of $\alpha$25 is below the instrumental noise level and could not be seen in the upper panel, the impact of confusion noise on the detector sensitivity may be fully reduced. For the other three populations, the sensitivity will be subject to the residuals of foreground at the most sensitive band, and the influence from the fiducial case would be relatively less significant than the q3 and $\alpha$5 cases. 
For the aASTROD-GW configuration, the residual will affect its sensitivity in a larger frequency range. As expected, the lowest residual is from the $\alpha$25 population, and the worst cases will be yielded by the q3 and $\alpha$5 models. And these residual levels are comparable to the residuals for eASTROD-GW in a frequency range of [0.1, 0.8] mHz comparing the curves in the upper plot.
The advantage of the aASTROD-GW would be a better sensitivity of frequencies higher than $\sim$0.8 mHz and lower than $\sim$0.1 mHz depending on different population models.

\section{Conclusions and discussions} \label{sec:conclusions}

In this work, we investigate the observations of galactic DWDs in the sub-mHz band. By employing two sensitivity configurations, eASTROD-GW and aASTROD-GW, we examine their detectability to four DWD populations as a succession of the LISA observation. The results show that the tens of thousands of binaries could be resolved by the sub-mHz mission(s) which accumulate around $\sim$1 mHz. The aASTROD-GW, with better sensitivity, can identify more faint DWDs than the eASTROD-GW. However, the resolvable binaries are a tiny fraction of the population, and GWs from the rest of the binaries will massively overlap and form the foreground in frequencies lower than $\sim$1 mHz. The sub-mHz mission will be subjected to the galactic foreground which could overwhelm the instrumental noises by orders. From another perspective, the galactic foreground would be a dominating stochastic GW signal for a sub-mHz mission, and their spectrum could be determined from the observation data. To examine the determination of these parameters from such a mission, the parameter inferences are performed for the foreground by using two mission configurations. The aASTROD-GW could better constrain the parameters of the confusion noise than the eASTROD-GW case since the former could observe the foreground in a larger frequency range and with better sensitivity. Furthermore, if the determined foreground could be subtracted from the data in an optimistic assumption, the PSD of foreground could be reduced by around two orders for the sub-mHz mission.
If this could be achievable, the sensitivity of detecting the primordial background might reach $10^{-14}$ critical cosmic closure density $\Omega_c$ \cite{Ni:2011ib}.

During our investigation, four galactic DWD populations are selected by setting a fixed binary fraction of 50\% in their formation. There are also alternative populations simulated by assuming the metallicity-dependent binary fraction, and the different setups result in different populations of galactic binaries \cite{Thiele:2021yyb}. As a result, the number of DWDs in the metallicity-dependent cases will be lower than in the fixed 50\% fraction, and the foreground is expected to be relatively lower than the results obtained in this study. We would deduce that the alternative foregrounds vary between the smallest population case ($\alpha$25) and the largest population case (q3). This study did not include the galactic binaries with neutron stars and/or black holes, and these more massive compact binaries, as well as their progenitors, would emit GW in sub-mHz frequencies \cite{Breivik:2019lmt,Wagg:2021cst,Sana:2012px,Chini:2012tw,2022RAA....22c5021Y}. The binary asteroids in the solar system would also be the potential sources for the sub-mHz missions \cite{Sullivan:2022uap}.
The galactic foregrounds could change with the size of binaries populations and the frequency distribution of the sources. More comprehensive studies would be required in the future.

The sub-mHz GW missions are expected to be planned after the LISA. The observation of the LISA, as well as the observations from the electromagnetic telescopes, will promote the understanding of the galactic population and binary evolution, and the succeeding sub-mHz mission(s) will get benefit from these precedent observations to characterize the foreground. On the other side, for the residual estimation, we optimistically subtract the inferred and restored foreground from the data considering it may overwhelming in the sensitive band of the detector. However, the galactic foreground may tangle with the instrumental noises, stochastic background, and other targeting sources. The global analysis, as a promising algorithm, may distinguish different GW sources simultaneously \cite{Littenberg:2020bxy,Littenberg:2023xpl}, and its development and application would be helpful to resolve the foreground in the sub-mHz band in the future.

\begin{acknowledgments}

G.W. was supported by the National Key R\&D Program of China under Grant No. 2021YFC2201903, and NSFC No. 12003059. 
Z.Y. was supported in part by the NSFC No. U1938114, and the Youth Innovation Promotion Association of CAS (id 2020265) and funds for key programs of the Shanghai Astronomical Observatory.
B.H. was supported in part by the National Key R\&D Program of China No. 2021YFC2203001.
W.T.N. was supported in part by the National Key R\&D Program of China No. 2021YFC2201901.
G. W. acknowledges Xiaobo Zou for helpful discussions. This work made use of the High Performance Computing Resource in the Core Facility for Advanced Research Computing at Shanghai Astronomical Observatory. The calculations in this work are performed by using the python packages $\mathsf{numpy}$ \cite{harris2020array}, $\mathsf{scipy}$ \cite{2020SciPy-NMeth} and $\mathsf{pandas}$ \cite{pandas}, and the plots are make by utilizing $\mathsf{matplotlib}$ \cite{Hunter:2007ouj}, $\mathsf{GetDist}$ \cite{Lewis:2019xzd} and $\mathsf{Component Library}$ \cite{ComponentLibrary}.

\end{acknowledgments}

\appendix

\section{GW Response formulation of TDI} \label{sec:appendix_response}

\begin{figure}[htb]
\includegraphics[width=0.46\textwidth]{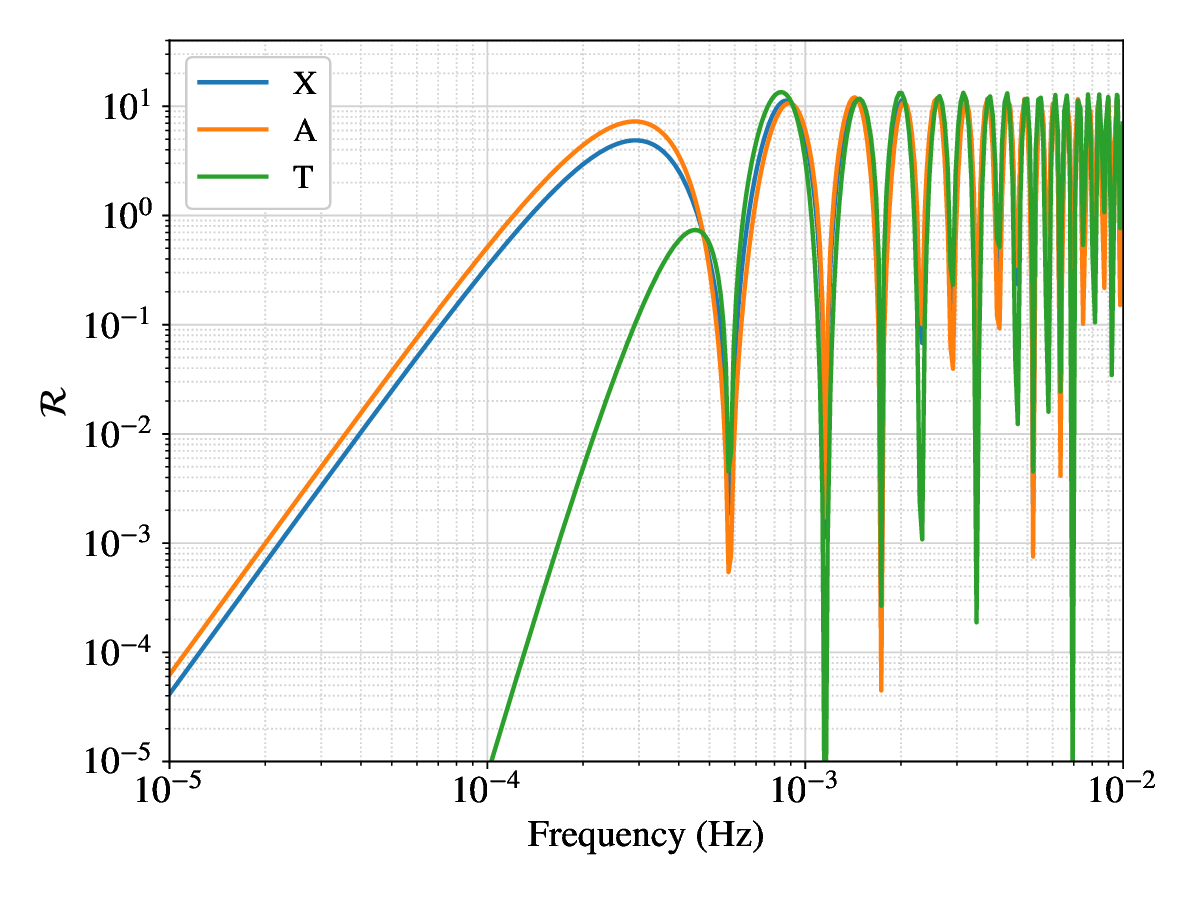}
\caption{The averaged GW response of the first-generation TDI Michelson X, A, and T from the ASTROD-GW. \label{fig:response}
}
\end{figure}

The GW propagation vector from a source locating at ecliptic longitude $\lambda$ and latitude $\theta$ (in the solar-system barycentric coordinates) will be
\begin{equation} \label{eq:source_vec}
 \hat{k}  = -( \cos \lambda \cos \theta, \sin \lambda \cos \theta ,  \sin \theta ).
\end{equation}
The $+$ and $\times$ polarization tensors of the GW signal with inclination angle $\iota$ of the source are
\begin{equation} \label{eq:polarizations-response}
\begin{aligned}
{\rm e}_{+} & \equiv \mathcal{O}_1 \cdot
\begin{pmatrix}
1 & 0 & 0 \\
0 & -1 & 0 \\
0 & 0 & 0
\end{pmatrix}
\cdot \mathcal{O}^T_1 \times \frac{1+\cos^2 \iota}{2} ,
\\
{\rm e}_{\times} &  \equiv \mathcal{O}_1 \cdot
\begin{pmatrix}
0 & 1 & 0\\
1 & 0 & 0 \\
0 & 0 & 0
\end{pmatrix}
\cdot \mathcal{O}^T_1 \times i (- \cos \iota ),
\end{aligned}
\end{equation}
with
\begin{widetext}
\begin{equation}
\mathcal{O}_1 =
\begin{pmatrix}
\sin \lambda \cos \psi - \cos \lambda \sin \theta \sin \psi & -\sin \lambda \sin \psi - \cos \lambda \sin \theta \cos \psi & -\cos \lambda \cos \theta  \\
     -\cos \lambda \cos \psi - \sin \lambda \sin \theta \sin \psi & \cos \lambda \sin \psi - \sin \lambda \sin \theta \cos \psi & -\sin \lambda \cos \theta  \\
         \cos \theta \sin \psi & \cos \theta \cos \psi & -\sin \theta
\end{pmatrix},
\end{equation}
where $\psi$ is polarization angle. The response to the GW in laser link from S/C$i$ to $j$ will be
\begin{equation} \label{eq:y_ij}
\begin{aligned}
y^{h}_{ij} (f) =&  \frac{ \sum_\mathrm{p} \hat{n}_{ij} \cdot {\mathrm{ e_p}} \cdot \hat{n}_{ij} }{2 (1 - \hat{n}_{ij} \cdot \hat{k} ) }
 \times \left[  \exp( 2 \pi i f (L_{ij} + \hat{k} \cdot p_i ) ) -  \exp( 2 \pi i f  \hat{k} \cdot p_j )  \right] ,
\end{aligned}
\end{equation}
\end{widetext}
where $\hat{n}_{ij}$ is the unit vector from S/C$i$ to $j$, $L_{ij}$ is the arm length from S/C$i$ to $j$, $p_i$ is the position of the S/C$i$ in the solar-system barycentric ecliptic coordinates. The GW response of a TDI channel is synthesized by the interferometric links, for instance, the GW response of the Michelson-X channel will be
\begin{equation}
\begin{aligned}
F^{h}_{ \rm X} (f) =& (-\Delta_{21} + \Delta_{21}  \Delta_{13}  \Delta_{31})  y^{h}_{12} \\
         & + (-1 + \Delta_{13}  \Delta_{31} )  y^{h}_{21} \\
         & + (\Delta_{31} - \Delta_{31}  \Delta_{12}  \Delta_{21})  y^{h}_{13} \\
         & + ( 1 - \Delta_{12}  \Delta_{21} )  y^{h}_{31}, 
\end{aligned}
\end{equation}
where $\Delta_{ij} = 2 \pi f L_{ij}$. The average GW response over all sky direction and polarization angle could be evaluated by following
\begin{equation}
\begin{aligned}
\mathcal{R}_{\rm TDI} (f) =& \frac{1}{4 \pi^2}  \int^{2 \pi}_{0} \int^{\frac{\pi}{2}}_{-\frac{\pi}{2}} \int^{\pi}_{0} |F^{h}_{ \rm TDI} (f, \iota=0)|^2 \cos \beta {\rm d} \psi {\rm d} \beta {\rm d} \lambda.
\end{aligned}
\end{equation}
The average GW response of the Michelson X, A, and T channels are shown in Fig. \ref{fig:response}. The average response of the E channel is identical to A. The response of the A or E channel is higher than X by a factor $\sqrt{3/2}$ in the lower frequencies \cite{Wang:1stTDI}. The T channel is much lower than other science channels in the lower frequency band, and the response becomes comparable to others in frequencies higher than $\sim$0.5 mHz. 

\section{Modulation of a monochromatic signal} \label{sec:modulation_waveform}

\begin{figure}[htb]
\includegraphics[width=0.48\textwidth]{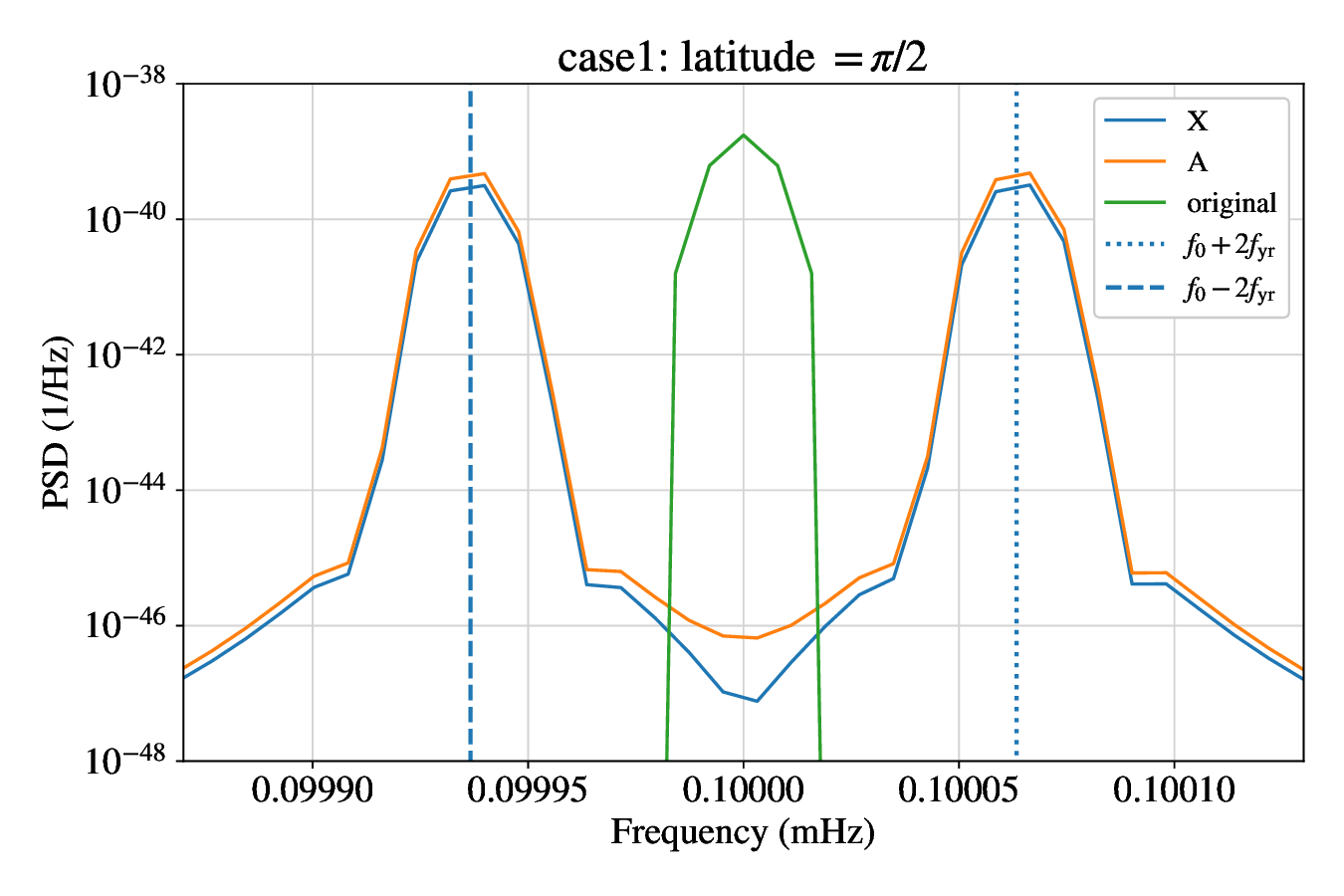}
\includegraphics[width=0.48\textwidth]{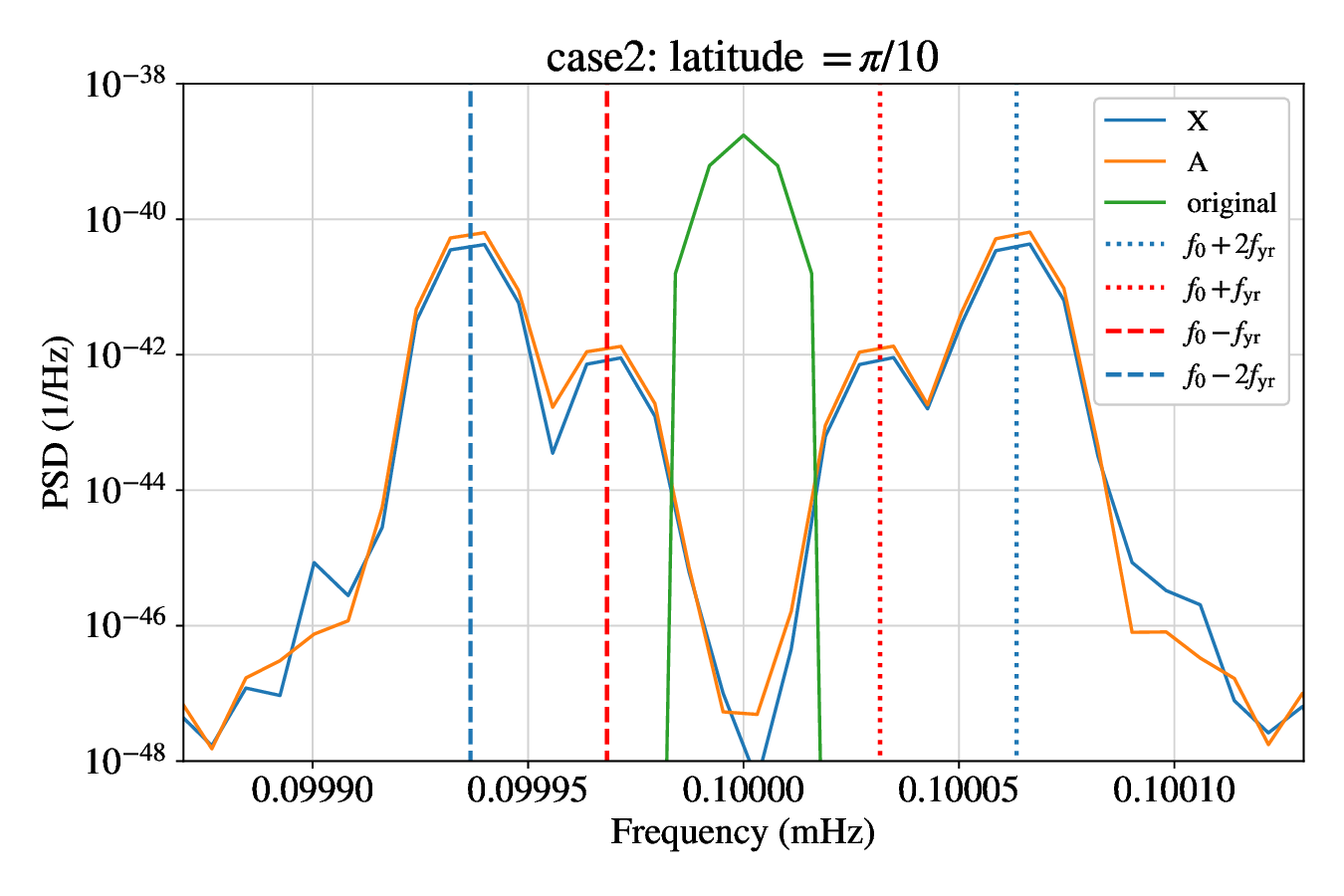}
\caption{The PSDs of monochromatic signals ($f_0 = 0.1$ mHz) observed by ASTROD-GW. The upper panel shows the results for polar case (ecliptic latitude $\theta = \pi/2$), while the lower panel shows the results for case2 ($\theta = \pi/10$). When the source is located in the polar direction, its GW frequency is shifted to $f_0 \pm 2 f_\mathrm{yr}$, where $f_\mathrm{yr} = 1/\mathrm{yr} \simeq 3\times 10^{-8}$ Hz, and it should be due to the rotation of interferometer. For case2, secondary modulated frequencies emerge at $f_0 \pm f_\mathrm{yr}$ which should be caused by the yearly motion of detector in the direction of source. \label{fig:modulation}
}
\end{figure}

For the ASTROD-GW and similar missions, the barycenter of the constellation will be close to the Sun, and the interferometer's yearly rotation will yield the modulated frequency for an observed signal. To examine the modulation effect, we numerically simulate the monochromatic signals from two directions: the first one is in the polar direction (ecliptic latitude $\theta=\pi/2$), and the other is inclined with respect to the ecliptic plane ($\theta=\pi/10$). The frequency of the monochromatic signal is set to $f_0 = 0.1$ mHz, and the amplitude is $A_0 = 6.9 \times 10^{-24}$. The PSDs of the observed signals over a four-year observation by ASTROD-GW are shown in Fig. \ref{fig:modulation}.

In the polar case, the observed frequencies in the TDI channels, X and A, are at $f_0 \pm 2 f_\mathrm{yr}$, where $f_\mathrm{yr} = 1/\mathrm{yr} \simeq 3\times 10^{-8}$ Hz, and this is due to the rotation of the interferometer and the symmetry of its antenna pattern. The antenna pattern of ASTROD-GW at $t$ and $t+\frac{1}{2} \mathrm{yr}$ would be the same. Compared to the stationary interferometer, the constant rotation would introduce a frequency shift of $\pm 2 f_\mathrm{yr}$. For case2, in addition to the frequency shift caused by the interferometer rotation, the relative motion in the source direction will cause secondary frequency modulation at $f_0 \pm f_\mathrm{yr}$. 

During a ten-year ASTROD-GW observation, the observed frequency of GW emitted by a DWD will modulate with the motion of detector and span across multiple frequency bins. However, in this investigation, the foreground spectra are smoothed by averaging over neighboring bins, as described in Section \ref{subsec:algorithm}. This smoothing process reduces the impact of modulation effects in the foreground evaluation, and our results will not be significantly affected by this modulation.

\nocite{*}
\bibliography{apsref}

\end{document}